\documentclass{JHEP3}

\JHEPspecialurl{http://jhep.sissa.it/JOURNAL/JHEP3.tar.gz}

\usepackage{multicol,bbm}
\usepackage{graphicx}
\usepackage[tight]{subfigure}
\usepackage{amsmath}
\usepackage{multirow}

\newcommand\fverb{\setbox\pippobox=\hbox\bgroup\verb}
\newcommand\fverbdo{\egroup\medskip\noindent%
			\fbox{\unhbox\pippobox}\ }
\newcommand\fverbit{\egroup\item[\fbox{\unhbox\pippobox}]}
\newbox\pippobox

\title{Sflavor mixing map viewed from a high scale
  in supersymmetric SU(5)}

\author{Pyungwon~Ko\\
        School of Physics, KIAS, Cheongnyangni-dong, Seoul, 130--722, Korea\\
	E-mail: \email{pko@kias.re.kr}}

\author{Jae-hyeon~Park\\
        INFN, Sezione di Padova, via F Marzolo 8, I--35131, Padova, Italy\\
        \email{jae-hyeon.park@pd.infn.it}}

\author{Masahiro~Yamaguchi\\
	Department of Physics, Tohoku University, Sendai 980--8578, Japan\\
	E-mail: \email{yama@tuhep.phys.tohoku.ac.jp}}

\preprint{{\tt KIAS-P07051, TU-814}\\\arXivid{0809.2784}}

\abstract{
  We study flavor violation in a supersymmetric SU(5) grand unification
  scenario in a model-independent way employing mass insertions.
  We examine how the quark and the lepton sector observables restrict
  sfermion mixings.
  With a low soft scalar mass,
  a lepton flavor violating process provides
  a stringent constraint
  on the flavor structure of right-handed down-type squarks.
  In particular, $\meg$ turns out to be highly susceptible to
  the 1--3 and 2--3 mixings thereof, due to
  the radiative correction from the top Yukawa coupling
  to the scalar mass terms of $\mathbf{10}$.
  With a higher scalar mass around the optimal value, in contrast,
  the quark sector inputs such as $B$-meson mixings and hadron
  electric dipole moment, essentially determine
  the room for sfermion mixing.
  We also discuss the recent deviation observed in $B_s$ mixing phase,
  projected sensitivity of forthcoming experiments,
  and ways to maintain the power of leptonic restrictions
  even after incorporating a
  solution to fix the incorrect quark--lepton mass relations.
}

\newcommand{\wt}{\widetilde}
\newcommand{\wh}{\widehat}

\newcommand{\order}{\mathcal{O}}

\newcommand{\GeV}{\ \mathrm{GeV}}
\newcommand{\MeV}{\ \mathrm{MeV}}

\newcommand{\MPl}{M_\mathrm{Pl}}
\newcommand{\Mgrav}{M_*}
\newcommand{\MGUT}{M_\mathrm{GUT}}
\newcommand{\MSUSY}{M_\mathrm{SUSY}}
\newcommand{\MR}{M_R}
\newcommand{\tb}{\tan\!\beta}
\newcommand{\cb}{\cos\!\beta}
\newcommand{\tmg}{\tau \rightarrow \mu \gamma}
\newcommand{\teg}{\tau \rightarrow e \gamma}
\newcommand{\meg}{\mu \rightarrow e \gamma}
\newcommand{\bsg}{B \rightarrow X_s \gamma}
\newcommand{\bdg}{B \rightarrow X_d \gamma}

\newcommand{\bdbdbar}{$B^0$--$\overline{B^0}$}
\newcommand{\bsbsbar}{$B_s$--$\overline{B_s}$}
\newcommand{\msl}{\wt{m}_{\wt{l}}}
\newcommand{\msd}{\wt{m}_{\wt{d}}}
\newcommand{\BsKsgam}{B_s \rightarrow K^* \gamma}
\newcommand{\ACPbsgam}{A_{CP}^{b \rightarrow s \gamma}}
\newcommand{\ACPbdgam}{A_{CP}^{b \rightarrow d \gamma}}
\newcommand{\ACPbsdgam}{A_{CP}^{b \rightarrow (s+d) \gamma}}
\newcommand{\ACPKstargam}{A_{CP}^{K^* \gamma}}
\newcommand{\SphiK}{S_{CP}^{\phi K}}
\newcommand{\SKstargam}{S_{CP}^{K^* \gamma}}
\newcommand{\Srhogam}{S_{CP}^{\rho \gamma}}
\newcommand{\SBsKsgam}{S_{CP}^{\BsKsgam}}

\newcommand{\ten}{T}
\newcommand{\fbar}{\overline{F}}
\newcommand{\adj}{\Sigma}
\newcommand{\fhiggs}{H}
\newcommand{\fbhiggs}{\overline{H}}
\newcommand{\NR}{N}
\newcommand{\lambdau}{\lambda_U}
\newcommand{\lambdad}{\lambda_D}
\newcommand{\lambdan}{\lambda_N}

\newcommand{\deltad}{\delta^d}
\newcommand{\deltal}{\delta^l}
\newcommand{\ded}[2]{(\deltad_{#1})_{#2}}
\newcommand{\del}[2]{(\deltal_{#1})_{#2}}

\newcommand{\incgr}[2][]{\IfFileExists{pr-tvmet/KNLO/HFAG/#2.pdf}{%
    \includegraphics[#1]{pr-tvmet/KNLO/HFAG/#2.pdf}}{%
    \includegraphics[#1]{#2.pdf}}}

\begin{document}

\section{Introduction}

The Large Hadron Collider (LHC) has started finally,
which we hope will be the first machine to produce supersymmetric particles
directly.
At this stage,
experimental input that is still playing a major role
in probing the soft supersymmetry breaking sector
and that will keep doing so even in the LHC era,
is the flavor changing neutral current (FCNC) and $CP$ violating processes.
 From this data, one can extract information on the
potential new sources of flavor and $CP$ violations
in the soft supersymmetry breaking terms
(see e.g.\ \cite{Gabbiani:1996hi} and papers that cite it).
A model of supersymmetry breaking/mediation, possibly in conjunction with
a model of flavor, should be compatible with this information.
In particular, the past two years have seen new measurements of
\bsbsbar\ mixing, both its size
\cite{DMsothers,Abulencia:2006ze}
and its phase \cite{Grossman:2006ce,phisothers,Abazov:2008fj}
(the latter still with low precision), which provide
new important restrictions on
the mixing between the second and the third families of down-type squarks
\cite{Ciuchini:2006dx,Endo:2006dm,gluinoloopBsothers,Ko:2008xb}.
On the other hand, a new experiment is going to explore the
lepton flavor violation (LFV) decay mode $\meg$,
squeezing its branching ratio down to the level of $10^{-13}$
\cite{MEG},
two orders of magnitude lower than the current upper bound.
Therefore, it can be regarded as timely to update an analysis
on supersymmetric flavor violation.

An interesting option in this style of model-independent analysis is
to work with a grand unified theory (GUT)\@.
We take the SU(5) group for example.
Since a single irreducible representation contains both quarks and leptons,
their flavor structures are related.
This enables us to use both quark sector and lepton sector processes
to look into a single source of flavor violation.
It is entertaining to see which observable is supplying a tighter constraint.
The outcome can serve as a hint concerning
which sector has a higher prospect
for discovery of FCNC mediated by sparticles.
For the scalar masses and trilinear couplings to obey the GUT symmetry,
the scale of supersymmetry breaking mediation should be higher than
the GUT scale.
We suppose that this scale $\Mgrav$ is given by
the reduced Planck scale
$\MPl / \sqrt{8\pi} \sim 2 \times 10^{18}\GeV$,
or very close to it, as is the case in a gravity mediation scenario.

This work is by no means the first attempt in this direction
\cite{Baek:2001kh,Hisano:2003bd,Ciuchini:2003rg,Cheung:2007pj,Ciuchini:2007ha,Borzumati:2007bn,Goto:2007ee}.
Most notably, there is a recent article that has performed
an
analysis in a similar framework \cite{Ciuchini:2007ha}.
Three differences are worth mentioning.
First,
we use the aforementioned $\meg$ decay mode to constrain the 1--3 and
the 2--3 mixings, in addition to $\teg$ and $\tmg$
which were considered in Ref.~\cite{Ciuchini:2007ha}.
This seemingly unrelated process becomes relevant,
and highly restrictive in some cases,
thanks to the radiative correction to the $\mathbf{10}$ representation
scalar mass matrix from the top Yukawa coupling and
the Cabibbo--Kobayashi--Maskawa (CKM) mixing \cite{Barbieri:1994pv}.
As a matter of fact, this mechanism has long been known and included
in many of the preceding model studies
\cite{Baek:2001kh,megtriplefromGUT,Calibbi:2006nq}.
Yet, this is the first instance of taking it into account
in a model independent analysis allowing for general flavor mixing
of sfermions, as far as we know.
Second, the authors of Ref.~\cite{Ciuchini:2007ha}
assume that the quark and the lepton mass eigenstates
at the GUT scale are aligned to a high degree.
This may or may not be the case if
a solution is incorporated for fixing the wrong quark--lepton mass relations.
Especially, the first and the second families are subject to
unlimited misalignment in general \cite{Baek:2001kh}.
We propose a method to overcome this obstacle to some extent.
Third, we elucidate the importance of the gaugino to scalar mass ratio
as a key parameter governing relative strengths of
the hadronic and the leptonic flavor violations.
We show expansions and shrinks of the territory ruled by
each of the two sectors.
In addition to these refinements,
we include remarks concerning the latest hint of anomaly
in the mixing phase of the $B_s$-meson \cite{Bona:2008jn,Barberio:2008fa}.

This paper is organized as follows.
In Section~\ref{sec:GUTFCNC},
we spell out basics of flavor physics in a supersymmetric SU(5) GUT model.
Section~\ref{sec:analysis} presents the procedure of numerical analysis
and the experimental inputs.
In Section~\ref{sec:results}, we exhibit the exclusion plot of
each mass insertion, and discuss how one can interpret
the plot conservatively
when the lagrangian has non-renormalizable terms
for accommodating the first and the second family fermion masses.
This section also has a collection of upper bounds on the sfermion mixings,
as well as deviations in selected
$CP$ asymmetries allowed by the other constraints.
With a summary, we conclude in Section~\ref{sec:conclusion}.
One can find notations of the soft supersymmetry breaking terms
and the mass insertion parameters in the appendix.

\section{SU(5) GUT and FCNC}
\label{sec:GUTFCNC}

\subsection{GUT relation between squark and slepton mixings}

Let us begin by reviewing basic elements of
a supersymmetric SU(5) grand unification model,
that are relevant to flavor physics.
The superpotential has
the Yukawa couplings and the right-handed neutrino mass terms,
\begin{equation}
  \label{eq:wgut}
  W_\mathrm{GUT} \supset
  - \frac{1}{4}\epsilon_{abcde} \lambdau^{ij}
  \ten_i^{ab} \ten_j^{cd} \fhiggs^e +
  \sqrt{2}\, \lambdad^{ij} \fbhiggs_a \ten_i^{ab} \fbar_{jb} -
  \lambdan^{ij} \NR_i \fbar_{ja} \fhiggs^a +
  \frac{1}{2} M_N^{ij} \NR_i \NR_j
  .
\end{equation}
Matter fields in $\mathbf{10}$ and $\overline{\mathbf{5}}$
representations are denoted by $\ten$ and $\fbar$, respectively,
$\mathbf{5}$ and $\overline{\mathbf{5}}$ Higgses by
$\fhiggs$ and $\fbhiggs$, respectively,
and a right-handed neutrino by $\NR$.
The indices $a, \ldots, e$ run over components of
the fundamental representation of SU(5), and
$i, j = 1,2,3$ indicate the family.
Obviously, $\lambdau$ and $M_N$ are symmetric matrices while
$\lambdad$ and $\lambdan$ are not.
The above Yukawa couplings, by themselves, predict mass unification
of down-type quarks and charged leptons at the GUT scale:
\begin{equation}
  m_e = m_d, \quad
  m_\mu = m_s, \quad
  m_\tau = m_b.
\end{equation}
Among these, the third relation is consistent with measurements
at low energies, while the first two are not.
One way to explain this discrepancy is to make corrections to
relatively smaller masses
by including the following non-renormalizable terms \cite{Ellis:1979fg}:
\begin{equation}
  \label{eq:wnr}
  \begin{aligned}
  W_\mathrm{NR} =& \
  \frac{1}{4} \epsilon_{abcde} \left(
    f_1^{ij} \ten_i^{ab} \ten_j^{cd} \frac{\adj^e_f}{\Mgrav} \fhiggs^f +
    f_2^{ij} \ten_i^{ab} \ten_j^{cf} \fhiggs^d \frac{\adj^e_f}{\Mgrav}
  \right) \\
  &+
  \sqrt{2} \left(
    h_1^{ij} \fbhiggs_a \frac{\adj^a_b}{\Mgrav} \ten_i^{bc} \fbar_{jc} +
    h_2^{ij} \fbhiggs_a \ten_i^{ab} \frac{\adj_b^c}{\Mgrav} \fbar_{jc}
  \right)
  +
  h_N^{ij} \NR_i \fbar_{ja} \frac{\adj^a_b}{\Mgrav} H^b
  ,
  \end{aligned}
\end{equation}
where $\adj$ is the adjoint Higgs multiplet
responsible for breaking SU(5) down to the Standard Model (SM) gauge group.
These terms will contribute to
the Yukawa couplings of the effective theory below the GUT scale,
expressed in terms of the SM fields as
\begin{equation}
  \label{eq:wssm}
    W_\mathrm{SSM} =
    Q^T Y_U \overline{U} H_u +
    Q^T Y_D \overline{D} H_d +
    L^T Y_E \overline{E} H_d +
    L^T Y_N N H_u +
    \frac{1}{2} N^T \! M_N N ,
\end{equation}
where the fields denoted by uppercase letters are
components of the GUT multiplets,
\begin{equation}
  \label{eq:tenfbar}
  \ten_i \simeq
  \{ Q, \overline{U}, \overline{E} \}_i, \quad
  \fbar_i \simeq
  \{ \overline{D}, L \}_i .
\end{equation}
The Yukawa couplings appearing in the superpotential
of~\eqref{eq:wssm} are related to those
in~\eqref{eq:wgut} and~\eqref{eq:wnr} by
\begin{subequations}
  \begin{align}
    \label{eq:Yulambdau}
    Y_U &= \lambdau + \xi \left( \frac{3}{5} f_1 +
    \frac{3}{20} f_2^S + \frac{1}{4} f_2^A \right) , \\
    Y_D &= \lambdad - \xi \left( \frac{3}{5} h_1 - \frac{2}{5} h_2 \right) ,\\
    Y_E^T &=
    \lambdad - \xi\, \frac{3}{5}\, ( h_1 + h_2 ) ,
    \\
    Y_N^T &= \lambdan + \xi\, \frac{3}{5}\, h_N,
  \end{align}
\end{subequations}
where the superscripts $S$ and $A$ denote
the symmetric and the antisymmetric part
of the given matrix, respectively.
The small number $\xi$ is defined by
\begin{equation}
  \xi \equiv 5 \frac{\sigma}{\Mgrav} \approx 10^{-2} ,
\end{equation}
where $\sigma$ is the vacuum expectation value (VEV) of $\adj$,
expressed as in
\begin{equation}
  \langle \adj \rangle = \sigma\, \mathrm{diag} (2,2,2,-3,-3) .
\end{equation}
The contribution from the non-renormalizable terms makes the difference,
\begin{equation}
  \label{eq:yukawadiff}
  Y_D - Y_E^T = \xi h_2 ,
\end{equation}
and this can account for
the first and the second family quark and lepton masses.

For this purpose,
Ref.~\cite{Ciuchini:2007ha} does not make use of the $\order(\xi)$ corrections,
but they rely on Georgi-Jarlskog mechanism \cite{Georgi:1979df}.
Their scenario corresponds to a case in our work
where the quark and the lepton mass eigenbases coincide, i.e.\
$U_L = U_R = \mathbf{1}$ in the formalism spelled out below.

Note that the proton lifetime depends on the structure of non-renormalizable
operators \cite{Emmanuel-Costa:2003pu}, thereby imposing a restriction on
the parameters appearing in~\eqref{eq:wnr}.
There are corners of the parameter space in conflict with
proton decay experiments.
The present work is not specific to a particular pattern of those terms
and is valid provided that they are Planck-suppressed.

In order to discuss flavor violation coming from the sfermion sector,
one should fix the basis of matter supermultiplets.
One can choose a basis of $\ten_i$ and $\fbar_i$ fields
such that
\begin{equation}
  \label{eq:initialyukawa}
  Y_U = V_Q^T \wh{Y}_U U_Q^*, \quad Y_D = \wh{Y}_D , \quad
  Y_E = U_L^T \wh{Y}_E U_R^*, \quad Y_N = U_L^T V_L^T \wh{Y}_N ,
\end{equation}
where the hat on a matrix signifies that the given matrix is diagonal
with positive elements \cite{moroi},
$V_Q$ and $V_L$ are unitary matrices in the
standard parametrization \cite{Chau:1984fp,PDG}
each with three mixing angles and one phase,
and $U_Q$, $U_L$, and $U_R$ are general unitary matrices.
Note that $Y_U$ may not be a symmetric matrix, unlike $\lambda_U$.
In this basis where $Y_D$ is diagonal,
$Y_E$ may not be diagonalized in general due to the
difference~\eqref{eq:yukawadiff},
and it should be decomposed into the above form
using $U_L$ and $U_R$.
These two unitary matrices describe the mismatch
between the down-type quark and the charged lepton mass eigenstates,
arising from breakdown of the Yukawa unification $Y_E^T = Y_D$
which is a consequence of SU(5) at the renormalizable level.
Since $U_L$ and $U_R$ are crucial in correlating
hadronic and leptonic processes, we need to examine their structures.
We can estimate
the size of an off-diagonal element of $Y_E$
in the unit of the tau Yukawa coupling,
\begin{equation}
  \frac{Y_E - \wh{Y}_D}{[\wh{Y}_E]_{33}} = 
  \frac{- \xi h_2^T}{m_\tau / (v \cb)}
  \approx - \cb \,h_2^T ,
\end{equation}
where $v \simeq 170\GeV$ is the Higgs VEV\@.
Notice the suppression by the factor $\cb$ for high $\tb$.
Assuming that each element of $h_2$ is not larger than $\order(1)$,
one can obtain approximate magnitudes of 1--3 and 2--3 mixings
\cite{Baek:2001kh},
\begin{equation}
  \label{eq:1323mixing}
\begin{aligned}
  {[U_L]_{3a}} &\approx - \cb\ [h_2]_{3a} , &
  {[U_L]_{a3}} &\approx \cb\ [U_L h_2^* U_R^\dagger]_{a3} , \\
  [U_R]_{3a}   &\approx - \cb\ [h_2^\dagger]_{a3} , &
  [U_R]_{a3}   &\approx \cb\ [U_L^* h_2^T U_R^T]_{3a} ,
\end{aligned}
\end{equation}
for $a = 1,2$.
Note that they are suppressed by $\cb$.
The other entries of $U_L$ and $U_R$ can be of $\order(1)$.
Finally, we relate the fields to the down-quark and
charged lepton mass eigenstates as
\begin{equation}
  \label{eq:embed}
  Q = q, \quad
  \overline{U} = U_Q^T \overline{u}, \quad
  \overline{E} = U_R^T \overline{e}, \quad
  \overline{D} = \overline{d}, \quad
  L = U_L^\dagger l .
\end{equation}
This leads us to the superpotential,
\begin{equation}
\begin{aligned}
    W_\mathrm{SSM} =& \
    q^T [ V_Q^T \wh{Y}_U ] \overline{u} H_u +
    q^T [ \wh{Y}_D ] \overline{d} H_d \\ &+
    l^T [ \wh{Y}_E ] \overline{e} H_d +
    l^T [ V_L^T \wh{Y}_N ] N H_u +
    \frac{1}{2} N^T \! M_N N .
\end{aligned}
\end{equation}
 One can notice that $V_Q$ is the CKM matrix at the GUT scale.
 If $M_N$ is diagonal in this basis,
 one also has $V_L = U_\mathrm{PMNS}^\dagger$.
 Otherwise, the lepton mixing matrix receives additional rotations
 for diagonalizing $M_N$.

Let us turn to the soft supersymmetry breaking sector.
The SU(5) symmetry relates the soft supersymmetry breaking terms of
squarks and sleptons in a single GUT multiplet.
The scalar mass terms are given by
\begin{equation}
  \label{eq:scalarmass}
  - \mathcal{L}_\mathrm{soft} \supset
  \fbar^\dagger m^2_{\fbar} \,\fbar +
  \ten^\dagger m^2_\ten \,\ten +
  \fbar^\dagger \frac{\adj}{\Mgrav} m^{2\prime}_{\fbar} \,\fbar +
  \ten^\dagger \frac{\adj}{\Mgrav} m^{2\prime}_\ten \,\ten +
  \cdots ,
\end{equation}
in which the higher dimensional terms involving $\adj$ are
suppressed by $\order(\xi)$.
In terms of these soft mass parameters of the GUT multiplets,
one can express the soft scalar mass matrices of the SM fields as
\begin{subequations}
  \label{eq:scalarmassrelations}
\begin{align}
  \label{eq:m2QUEm2T}
    m^2_Q &= m^2_{\ten} + \frac{1}{10} \xi\, m^{2\prime}_{\ten}, \quad
    m^{2*}_U = m^2_{\ten} - \frac{2}{5} \xi\, m^{2\prime}_{\ten}, \quad
    m^{2*}_E = m^2_{\ten} + \frac{3}{5} \xi\, m^{2\prime}_{\ten}, \\
    m^{2*}_D &= m^2_{\fbar} + \frac{2}{5} \xi\, m^{2\prime}_{\fbar}, \quad
    m^2_L = m^2_{\fbar} - \frac{3}{5} \xi\, m^{2\prime}_{\fbar},
\end{align}
\end{subequations}
using~\eqref{eq:tenfbar}.
 From these expressions and~\eqref{eq:embed}, one can see
that the mass insertion parameters
of down-type squarks and sleptons at the GUT scale are linked by
\begin{subequations}
\label{eq:mirelations}
\begin{align}
\label{eq:mirelationRR}
  \deltal_{LL} &= U_L\,\delta^{d*}_{RR}\, U_L^\dagger
  + \order(\xi) ,
  \\
\label{eq:mirelationLL}
  \deltal_{RR} &= U_R\,\delta^{d*}_{LL}\, U_R^\dagger
  + \order(\xi) .
\end{align}
\end{subequations}
We can notice two possible sources of deviation from the naive equalities
\cite{Ciuchini:2003rg},
\begin{equation}
  \label{eq:naivecorr}
  \deltal_{LL} = \delta^{d*}_{RR}, \quad
  \deltal_{RR} = \delta^{d*}_{LL} .
\end{equation}
One is the higher dimensional terms in~\eqref{eq:scalarmass},
which makes the $\order(\xi)$ corrections,
and the other is $U_L$ and $U_R$, the unitary transformations
parametrizing the misalignment
between the down-type quark and the charged lepton mass eigenstates.
The former type of corrections is negligible compared to the typical size
of a scanning mass insertion parameter appearing later on.
On the other hand, these corrections might be comparable to
the renormalization group (RG) contribution to $\deltal_{RR}$.
Unless they are tuned in such a way that they cancel out
the RG-generated $\deltal_{RR}$,
they nevertheless do not undermine the importance of $\meg$ constraint.
The latter needs more consideration.
Obviously,
$U_L$ and $U_R$ depend on $h_2$ through $Y_E$.
If $h_2$ is diagonal in the basis where $Y_D$ is diagonal,
$U_L$ and $U_R$ are unit matrices, and~\eqref{eq:naivecorr}
becomes a fairly good approximation correlating
squark and slepton flavor mixings.
If $h_2$ is not diagonal, the correlation gets loose,
but in many cases, LFV processes can still
give meaningful restrictions on the down-type squark mixings,
thanks to the suppression of 1--3 and 2--3 mixings
shown in~\eqref{eq:1323mixing}.
Examples of this situation will be presented in
Section~\ref{sec:nrlfv}.

In a similar way, the GUT symmetry links
the scalar trilinear coupling terms of squarks and sleptons
so that their chirality-flipping mass insertions have the relations,
\begin{equation}
  \deltal_{LR} = U_L\,\delta^{d\,T}_{LR}\,U_R^\dagger +
  \order(\xi) \times A_0 \langle H_d \rangle / \msl^2 ,
\end{equation}
where $A_0$ is the overall scale of the $A$-terms and
$\msl$ is the average slepton mass.
In what follows, we do not use this expression
since we will ignore the $A$-term contributions
to flavor violating processes.

\subsection{RG running of scalar masses}
\label{sec:running}

RG running from one scale down to a lower scale
generates off-diagonal elements of a scalar mass matrix.
For our purpose, we need to consider two intervals of scale:
from $\Mgrav$ to $\MGUT$, and from $\MGUT$ (via $\MR$) to $\MSUSY$.
The former is needed to determine the boundary condition
to give on the soft supersymmetry breaking terms at the GUT scale,
and the latter is to connect the given boundary condition
with low energy observables.

First, we think of running between $\Mgrav$ and $\MGUT$.
Using one-loop approximation,
the RG-induced off-diagonal elements can be written as
\cite{Hall:1985dx,RGE10and5bar}
\begin{subequations}
  \begin{align}
    \label{eq:Deltagm2T}
    \Delta_\mathrm{g} m^2_{\ten} \simeq
    & - \frac{2}{(4\pi)^2} [
    3 \lambdau^* \lambdau^T + 2 \lambdad^* \lambdad^T
    ](3 m^2_0 + |A_0|^2)\ln \frac{\Mgrav}{\MGUT} ,
    \\
    \label{eq:Deltagm2F}
    \Delta_\mathrm{g} m^2_{\fbar} \simeq
    & - \frac{2}{(4\pi)^2} [
    4 \lambdad^\dagger \lambdad +
    \lambdan^\dagger \lambdan
    ](3 m^2_0 + |A_0|^2)\ln \frac{\Mgrav}{\MGUT} ,
  \end{align}
\end{subequations}
where $m_0$ is the scalar mass and $A_0$ is the trilinear scalar coupling.
Let us focus on the mass matrix of $T$ fields, which feeds into
the mixings of left-handed squarks and right-handed sleptons.
 From~\eqref{eq:Yulambdau}, \eqref{eq:initialyukawa},
\eqref{eq:m2QUEm2T},
and \eqref{eq:Deltagm2T},
one can obtain the following form of RG contribution to
the $LL$ squark mixing at the GUT scale,
\begin{equation}
  \label{eq:dedLLRG}
  \ded{ij}{LL} \simeq
  - \frac{6}{(4\pi)^2}
  \,[V_Q^\dagger \wh{Y}_U^2 V_Q]_{ij}\,
  \frac{3 m^2_0 + |A_0|^2}{\msd^2\,(\MGUT)}
  \ln \frac{\Mgrav}{\MGUT} + \order(\xi) .
\end{equation}
The $\order(\xi)$ correction in the second term is not necessarily
smaller than the first term coming from the CKM mixing and 
the large top quark Yukawa coupling.
Neither is it very likely, however, that they cancel out leading to a value
much smaller than the first term.
That is, the left-handed squark mixing in the above expression,
without the $\order(\xi)$ correction,
can be regarded as the minimal value of $\ded{ij}{LL}$
that is expected in a supersymmetric SU(5) model
with the cutoff at $\Mgrav$.
Let us record the CKM matrix dependence of the above minimal mass insertions,
\begin{equation}
  \label{eq:dedLLCKM}
  \ded{12}{LL} \sim V_{td}^* V_{ts} \sim \lambda^5, \quad
  \ded{13}{LL} \sim V_{td}^* V_{tb} \sim \lambda^3, \quad
  \ded{23}{LL} \sim V_{ts}^* V_{tb} \sim \lambda^2,
\end{equation}
where we also express them as powers of $\lambda$, sine of the Cabibbo angle.

Using~\eqref{eq:mirelationLL},
one can get the right-handed slepton mixing
from~\eqref{eq:dedLLRG}.
Again, we drop the $\order(\xi)$ term in~\eqref{eq:mirelationLL},
assuming that it does not conspire with the first term to result in
a drastic cancellation.
If $U_R$ is an identity matrix, $\del{ij}{RR}$ has the same pattern
as~\eqref{eq:dedLLCKM}.
Otherwise, one should take the misalignment into account.
As~\eqref{eq:1323mixing} shows that
the 1--3 and 2--3 mixings are suppressed,
one can rephrase~\eqref{eq:mirelationLL} into
\begin{equation}
  \label{eq:dela3RR}
  \del{a3}{RR} =
  [U_R]_{ab}\, \ded{b3}{LL}^* \, [U_R]_{33}^* +
  \order(\cos^2\!\beta\, \deltad_{LL}) , \quad
  a, b = 1, 2 ,
\end{equation}
where $[U_R]_{ab}$, the upper-left $2 \times 2$ submatrix of $U_R$,
is approximately unitary.
[Supposing universal scalar masses at $\Mgrav$,
one actually has
another term of the form $[U_R]_{a3}\, \ded{33}{LL} \, [U_R]_{33}^*$
where (with an abuse of notation
what we here call) $\ded{33}{LL}$
is given by setting $i = j = 3$ in~(\ref{eq:dedLLRG}).
In what follows we discard this term although it
can be larger than what is kept in the above equation.
Even if it happens to be non-negligible,
it generically enlarges the rate of $\meg$,
only to reinforce the sensitivity of this LFV channel.]
Keeping only the powers of $\lambda$, one can schematically rewrite this as
\begin{equation}
  \label{eq:delRRCKM}
  \del{13}{RR} \sim [U_R]_{11} \lambda^3 + [U_R]_{12} \lambda^2,
  \quad
  \del{23}{RR} \sim [U_R]_{21} \lambda^3 + [U_R]_{22} \lambda^2.
\end{equation}
The mixing between the first and the second families,
described by $[U_R]_{ab}$, is not particularly restricted to be small.
There can be small, large, or no mixing.
One finds that $\del{a3}{RR}$ is generically
not much smaller than $\lambda^3$,
unless the mixing is fine-tuned in such a way that
the two terms cancel out in either of~\eqref{eq:delRRCKM}.
For example, the mixing angle should be tuned between
$- \lambda \pm \lambda^2$
in order to have $|\del{13}{RR}| \lesssim \lambda^4$.

Next, we should turn to the running below $\MGUT$.
Before examining an off-diagonal entry of a scalar mass matrix,
let us recall the running of a diagonal element
since a mass insertion parameter is normalized by it.
Squark and slepton masses at $\MSUSY$
are approximately related to the GUT scale variables by
\begin{subequations}
\begin{align}
  \label{eq:squarkdiagonalrunning}
  \msd^2\, (\MSUSY) &\approx 
  (1 + 6 x)\, m^2_0 , \\
  \label{eq:sleptondiagonalrunning}
  \msl^2\, (\MSUSY) &\approx m^2_0 ,
\end{align}
\end{subequations}
with the definition of gaugino to scalar (squared) mass ratio,
\begin{equation}
  \label{eq:x}
  x \equiv M_{1/2}^2/m^2_0 .
\end{equation}
The squark mass increases considerably by the gaugino mass contribution.
The slepton mass actually receives
a small correction from the gaugino mass, but it can be ignored
for later discussions.
These facts will be crucial to understanding parameter dependence
of a constraint.

Unless $\tb$ is extremely high,
an off-diagonal element of $m^2_D$ does not run significantly,
while running of the left-handed squark mass matrix makes the difference
\cite{AlvarezGaume:1981wy},
\begin{equation}
  \label{eq:m2QRG}
  \Delta_\mathrm{s} [m^2_Q]_{ij} \simeq
  - \frac{2}{(4\pi)^2}
  \,[V_Q^\dagger \wh{Y}_U^2 V_Q]_{ij}\,
  (3 m^2_0 + |A_0|^2)
  \ln \frac{\MGUT}{\MSUSY} .
\end{equation}
Using these facts and~\eqref{eq:squarkdiagonalrunning},
we can associate squark mass insertions at $\MSUSY$ to those at $\MGUT$ as
\begin{subequations}
  \label{eq:squarkMIdilution}
\begin{align}
  \ded{ij}{RR} (\MSUSY) &\approx \frac{\ded{ij}{RR} (\MGUT)}{1 + 6 x} , \\
  \label{eq:squarkdLLdilution}
  \ded{ij}{LL} (\MSUSY) &\approx
  \frac{\ded{ij}{LL} (\MGUT) + q_{ij}}{1 + 6 x} ,
\end{align}
\end{subequations}
with the definition
\begin{equation}
  \label{eq:qij}
  q_{ij} \equiv \Delta_\mathrm{s} [m^2_Q]_{ij} / m^2_0 .
\end{equation}

In a parallel way, one can relate slepton mass insertions
at a low scale to those at a high scale by
\begin{subequations}
\label{eq:delLLRRRG}
\begin{align}
  \del{ij}{RR} (\MSUSY) &\approx \del{ij}{RR} (\MGUT) , \\
  \label{eq:delLLRG}
  \del{ij}{LL} (\MSUSY) &\approx \del{ij}{LL} (\MGUT) + l_{ij} ,
\end{align}
\end{subequations}
using~\eqref{eq:sleptondiagonalrunning} and the definition
$l_{ij} \equiv \Delta_\mathrm{s} [m^2_l]_{ij} / m_0^2$
with the radiative correction to the off-diagonal slepton mass matrix entries
\cite{Borzumati:1986qx},
\begin{equation}
  \label{eq:m2LRG}
  \Delta_\mathrm{s} [m^2_l]_{ij} \simeq
  - \frac{2}{(4\pi)^2}
  \,[V_L^\dagger \wh{Y}_N^2 V_L]_{ij}\,
  (3 m^2_0 + |A_0|^2)
  \ln \frac{\MGUT}{\MR} .
\end{equation}
This estimate is based on the assumption that the right-handed neutrinos
are degenerate so that they are integrated out at a single scale $\MR$.
If they are not degenerate, it is modified to involve
mixings, phases, and eigenvalues of $M_N$ (see e.g.\ \cite{Hisano:2003bd}).
Even in this case, it has been shown that one can use
the above form of expression by replacing $\MR$
with the largest eigenvalue of $M_N$,
if there is a large hierarchy among the right-handed
neutrino masses \cite{Hisano:2003bd}.
Unlike the quark sector,
we do not yet have much information on the neutrino Yukawa couplings.
They can be of $\order(1)$ in the case of heavy right-handed neutrinos,
or extremely small if the neutrino masses are of Dirac type.
Even if we suppose that seesaw mechanism is working,
a vast range of right-handed neutrino mass scale is possible,
from around the GUT scale down to the weak scale.
Although the lepton mixing angles have been measured to an extent,
they cannot be directly related to the mixing matrix $V_L$
due to the additional degrees of freedom in $M_N$,
the Majorana right-handed neutrino mass matrix.
Moreover, the hierarchy of neutrino masses is unknown yet.
As the magnitude of $l_{ij}$ in one model
can greatly differ from another,
we choose to drop it in the following analysis.
Therefore, the results shown later are legitimate only for a scenario
where right-handed neutrinos are light enough for $l_{ij}$ to be
negligible in~\eqref{eq:delLLRG}.
(For a study on a case with a large neutrino Yukawa coupling
and a specific boundary condition on the soft terms,
see e.g.\ \cite{Goto:2007ee,Dutta:2008xg,Hisano:2008df}.)

Nevertheless, there are circumstances where one can tell consequences of
non-negligible $l_{ij}$.
Here, we assume that $U_L$ is a unit matrix.
This assumption will be relaxed in Section~\ref{sec:nrlfv}.
If neutrino Yukawa couplings are large, they affect not only
the running below, but also above $\MGUT$, of $m^2_l$.
Thus, $\del{ij}{LL} (\MGUT)$ is decomposed into two pieces,
\begin{equation}
  \label{eq:delGUTdelMPl}
  \del{ij}{LL} (\MGUT) \approx \del{ij}{LL} (\Mgrav) + \alpha l_{ij} ,
\end{equation}
where the first term represents possible flavor non-universality
at the reduced Planck scale,
and the second is the RG contribution with
\begin{equation}
  \label{eq:alpha}
  \alpha \equiv \frac{\ln (\Mgrav/\MGUT)}{\ln (\MGUT/\MR)} .
\end{equation}
What we will do in the following sections
is to search for a set of viable values
of $\del{ij}{LL} (\MGUT)$, imposing experimental constraints.
In terms of the variables in~\eqref{eq:delGUTdelMPl},
we can interpret this procedure in two different ways:
we fix $l_{ij}$ and scan over $\del{ij}{LL} (\Mgrav)$, or the other way around.
As an example of the first option,
suppose that one studies a neutrino mass model in which
the neutrino Yukawa matrix is given, but there is a room for flavor
mixing in the soft supersymmetry breaking terms.
In this case, one can easily guess
the allowed region of
$\del{ij}{LL} (\MGUT) = \ded{ij}{RR}^* (\MGUT)$ from
the one shown in Section~\ref{sec:regions}
using~\eqref{eq:delLLRG}: shift the region by $- l_{ij}$.
This method is applicable to a model with
non-degenerate right-handed neutrinos as well.
Regarding the second option,
one can imagine a situation where the only source of $\fbar$ mixing
is the neutrino Yukawa matrix, i.e.\
$\del{ij}{LL} (\Mgrav) = 0$.
Under this condition, \eqref{eq:delLLRG} can be rewritten as
\begin{equation}
  \label{eq:delMGUTdelMSUSY}
  \del{ij}{LL} (\MGUT) \approx
  \frac{\alpha}{1 + \alpha}\,\del{ij}{LL} (\MSUSY) ,
\end{equation}
which relies on the degeneracy of right-handed neutrinos.
Obviously, the allowed region of $\del{ij}{LL} (\MGUT)$ is given by
shrinking the one in Section~\ref{sec:regions}
by the factor $\alpha/(1 + \alpha)$.

In this subsection, we used one-loop estimates to understand
the qualitative behaviors of squark and slepton mixings,
but we numerically solve RG equations
for quantitative analysis in the subsequent sections.

\section{How to impose constraints on scalar mixings}
\label{sec:analysis}

\subsection{Scheme}

One popular way to constrain sfermion mixings in a model-independent
fashion is to scan over one mass insertion parameter at a time, while
setting the other parameters to zero.  The practical reason to assume all
but one of the parameters to be zero is that it is difficult or impossible to
take more than one complex mass insertions as free variables and plot
the allowed volume.  Despite its makeshift motive, this
strategy works as long as the parameter being swept by itself makes
the dominant contribution to the process in consideration.
However, there are cases where presence of another mass insertion
amplifies the contribution from the scanned parameter,
thereby rendering the constraint from a process much tighter.

\begin{figure}
  \centering
\subfigure[single insertion]{\incgr{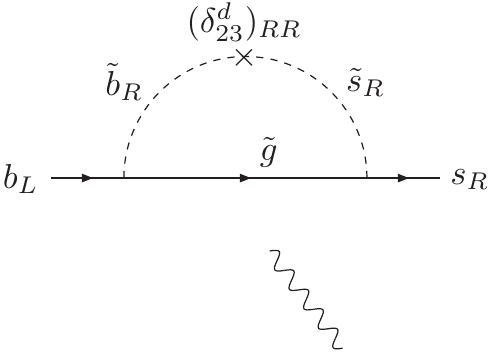}}
\subfigure[double insertion]{\incgr{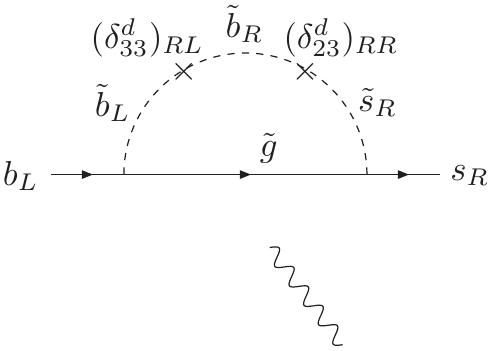}}
  \caption{Gluino loop contributions of $\ded{23}{RR}$ to $\bsg$.}
  \label{fig:bsg}
\end{figure}

A well known example is $\bsg$.
For instance, a single $\ded{23}{RR}$ insertion contributes to this decay
via the gluino loop shown in Fig.~\ref{fig:bsg}~(a).
If one takes into account nonzero $\ded{33}{RL}$ insertion as well,
the diagram in Fig.~\ref{fig:bsg}~(b) with double insertions
can make an additional contribution \cite{Ko:2008xb,Gabbiani:1988rb},
whose amplitude is enhanced by $\tb$
relative to the single insertion graph due to the chirality flip on the
gluino propagator.
The reason for including the double insertion diagram, namely
considering nonzero $\ded{33}{RL}$ in addition to
the $\ded{23}{RR}$ under inspection, is not only that
it can significantly increase the $\bsg$ branching ratio,
but also that $\ded{33}{RL} \equiv m_b (A - \mu \tb) / \msd^2$
is generically present and therefore it should not be ignored.
An $s \rightarrow d$ equivalent has been used in the study of
$\epsilon'/\epsilon_K$ \cite{epsp}.

Another example is \bsbsbar\ mixing.
This process is affected by $\ded{23}{RR}$ as well.
However, the \bsbsbar\ mixing constraint on $\ded{23}{RR}$
greatly depends on the size of $\ded{23}{LL}$ \cite{Gabbiani:1988rb},
and therefore it matters what value of the $LL$ insertion we choose
when we are focusing on the $RR$ mixing.
Apart from the simple-minded choice of vanishing $LL$ insertion,
one option is to set $\ded{23}{LL} = \ded{23}{RR}$ \cite{Ciuchini:2006dx},
which may be expected from a left-right symmetry.
Another well-motivated value of $\ded{23}{LL}$ is the one generated by
RG running from the scale where the boundary condition is given down to
the sparticle mass scale \cite{Endo:2006dm}.
This value is shown in a rather obscure form in~\eqref{eq:squarkdLLdilution}
and \eqref{eq:dedLLRG}.
It comes from the CKM mixing of quark Yukawa couplings
and is expected even with universal soft supersymmetry breaking
terms at $\Mgrav$.
It should be reasonable to expect
at least this amount of $LL$ insertion,
even if one allows for general non-universal boundary condition,
which is the case in this work.

\FIGURE{%
  \incgr{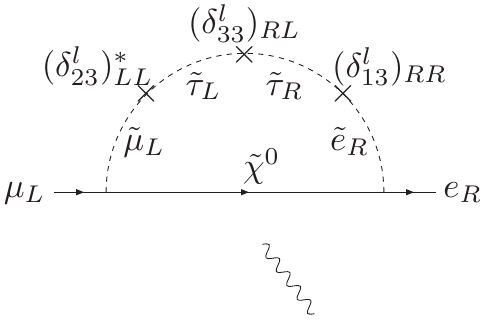}
  \caption{Neutralino loop contribution to $\meg$ with triple mass insertions.}
  \label{fig:megtriple}
}%
In the framework of supersymmetric GUT, the story can be extended
in a more interesting way.
The aforementioned parameter $\ded{23}{RR}$
is related to
$\del{23}{LL}$ at the GUT scale, and it can lead to LFV\@.
An obvious decay mode is $\tmg$
\cite{Hisano:2003bd,Ciuchini:2003rg,Cheung:2007pj,Ciuchini:2007ha}.
It can serve as another constraint on $\ded{23}{RR}$,
under the assumption of SU(5) grand unification.
A less obvious mode is $\meg$.
Due to the GUT symmetry,
the CKM mixing leads to an off-diagonal element of the scalar mass matrix
of the entire $\mathbf{10}$ members,
while they run from the reduced Planck scale down to the GUT scale.
With the help of $\del{13}{RR}$ produced in this way,
one can complete a diagram for $\meg$ with triple mass insertions
shown in Fig.~\ref{fig:megtriple}.
This diagram receives $m_\tau / m_\mu$ enhancement relative to
the usual chargino loop since it is proportional to $\del{33}{RL}$
\cite{Baek:2001kh,megtriplefromGUT,Hisano:1995cp,Paradisi:2005fk}.
Therefore it can give a strong restriction on $\del{23}{LL}$ and thus on
$\ded{23}{RR}$.

The above examples illustrate how much a constraint on a given
mass insertion parameter can be strengthened due to the presence of
another insertion.
Then, the question would be what the reasonable default value of
a mass matrix element is,
while a particular mass insertion is being scanned.
In this work, we take the following scheme for choosing the default value
of a soft supersymmetry breaking parameter:
by default,
the $\mathbf{10}$ soft scalar mass matrix elements
are set to the RG-induced values from the top Yukawa coupling and
the CKM mixing, and the off-diagonal components of
$\overline{\mathbf{5}}$ mass matrix
are set to zero;
we ignore scalar trilinear couplings
supposing that a loop graph arising from a nontrivial $A$-term
does not accidently cancel the contributions considered later.

Using this scheme, we carry out a numerical analysis
taking the following steps.
 From the Yukawa couplings and gauge couplings at the weak scale,
those at the GUT scale are computed by solving
the one-loop RG equations.
In this process, neutrino Yukawa couplings are ignored.
After reaching the GUT scale,
we move to the basis of $q$, $\overline{u}$, $\overline{d}$,
$l$, and $\overline{e}$ such that
$Y_d$ and $Y_e$ are diagonal.
We assume that $U_L$ and $U_R$ are identity matrices,
and thus $q$, $\overline{d}$, $l$, $\overline{e}$ are identical
to their uppercase counterparts in~\eqref{eq:embed}.
In terms of the superpotential parameters,
this corresponds to the case where $h_2$ in~\eqref{eq:wnr} is such that
it reproduces the observed down-type quark and charged lepton masses,
and is diagonal in the basis where $Y_D$ is diagonal.
Consequences of relaxing this assumption will be discussed in
Section.~\ref{sec:nrlfv}.
In this super-CKM basis of
down-type quarks and charged leptons at the GUT scale,
we set the soft mass matrix of squarks to the form,
\begin{equation}
  \label{eq:massmatrix}
  m^2_q = m^2_0
  \left(
    \begin{array}{ccc}
      1 & \ded{12}{LL} & \ded{13}{LL} \\
      \ded{12}{LL}^* & 1 & \ded{23}{LL} \\
      \ded{13}{LL}^* & \ded{23}{LL}^* & 1
    \end{array}
  \right) , \quad
  m^2_d = m^2_0
  \left(
    \begin{array}{ccc}
      1 & 0 & \ded{13}{RR} \\
      0 & 1 & \ded{23}{RR} \\
      \ded{13}{RR}^* & \ded{23}{RR}^* & 1
    \end{array}
  \right) ,
\end{equation}
and we determine the slepton soft masses using~\eqref{eq:scalarmassrelations}
neglecting the $\order (\xi)$ corrections.
The other scalar masses including those of Higgses are universally
put to $m_0$.
The trilinear scalar couplings are set to zero.
With these boundary conditions given at the GUT scale,
the one-loop RG evolution
of the lagrangian parameters is performed
down to the weak scale.
In order to fill out the mass matrices of scalars, charginos, and neutralinos,
we determine $\mu$ from the electroweak symmetry breaking condition,
choosing the positive sign.
We have numerically checked that changing the sign of $\mu$ does not make
a substantial difference.
Then, we have all the sparticle mass matrices needed to calculate
flavor and $CP$ violation quantities.
We do not use mass insertion approximation, but
employ mass eigenvalues and mixing matrices, thereby
taking account of multiple insertion graphs automatically.
For a quark sector amplitude, we keep only gluino loops, and disregard
parametrically suppressed corrections from
neutralino, chargino, and charged Higgs exchanges.

Regarding patterns of the mass insertion parameters in~\eqref{eq:massmatrix},
we consider the four cases displayed in Table~\ref{tab:insertions}.
A parameter indicated as `free' is a variable to be scanned over,
and the other three are fixed at the respective specified numbers,
according to the policy outlined above.
Those numbers have been obtained by
solving the RG equations
for the soft scalar mass matrices
with universal boundary conditions at the reduced Planck scale
down to the GUT scale
in a supersymmetric SU(5) model with minimal field content
\cite{Baek:2001kh}.
In this procedure, we have ignored
effects of non-renormalizable operators on the running of
scalar mass matrices.
The size of $\ded{ij}{LL}$ depends on $m_0$, $M_{1/2}$, and $\tb$,
where $M_{1/2}$ is the unified gaugino mass at $\MGUT$.
This dependence is taken into account in a plot
for a different set of input parameters, although
the change from the value shown in the table is insignificant.
\TABLE{%
  \renewcommand{\arraystretch}{1.1}
  \begin{tabular}{cccccc}
    \hline
    Fig. & $|\ded{12}{LL}|$ &
    $|\ded{13}{LL}|$ & $|\ded{23}{LL}|$ & $|\ded{13}{RR}|$ & $|\ded{23}{RR}|$\\
    \ref{fig:23RR} & $4.8 \times 10^{-5}$ &
    $1.5 \times 10^{-3}$ & $7.4 \times 10^{-3}$ & 0 & free            \\
    \ref{fig:13RR} & $4.8 \times 10^{-5}$ &
    $1.5 \times 10^{-3}$ & $7.4 \times 10^{-3}$ & free & 0               \\
    \ref{fig:23LL} & $4.8 \times 10^{-5}$ &
    $1.5 \times 10^{-3}$ & free             & 0 & 0               \\
    \ref{fig:13LL} & $4.8 \times 10^{-5}$ &
    free             & $7.4 \times 10^{-3}$ & 0 & 0               \\
    \hline
  \end{tabular}
  \caption{Values of mass insertion parameters to be given as
    boundary conditions at the GUT scale,
    for the case with $m_0 = 220\GeV$, $M_{1/2} = 180\GeV$,
    and $\tb = 5$.
    The phase of a fixed $\ded{ij}{LL}$ is equal to
    $\arg( - V_{ti}^* V_{tj} )$,
    as can be expected from~\eqref{eq:dedLLRG}.
    The first column points to the plot of each free variable.}
  \label{tab:insertions}
}%

\subsection{Observables}

We summarize observables from the quark sector
and how we use them as constraints,
in Table~\ref{tab:hconstraints}.
\TABLE[b]{
  \renewcommand{\arraystretch}{1.1}
  \begin{tabular}{ccc}
    \hline
    Observable       & Measured value & Imposed constraint \\
    $\Delta M_{B_d}$ & $0.507 \pm 0.004\ \mathrm{ps}^{-1}$
    \cite{Barberio:2008fa} & $0.507\ \mathrm{ps}^{-1} \pm 30\%$ \\
    $\sin 2\beta$    & $0.681 \pm 0.025$ \cite{Barberio:2008fa} &
    $2\ \sigma$ \\
    $\cos 2\beta$    & $> -0.4$ \cite{Bona:2006ah} \\
    $B(\bdg)$        &
    $(3.1 \pm 0.9^{+0.6}_{-0.5} \pm 0.5) \times 10^{-6}$ \cite{LP07nakao}
    & $[5 \times 10^{-7}, 10^{-5}]$ \\
    $\Delta M_{B_s}$ & $17.77 \pm 0.10 \pm 0.07\ \mathrm{ps}^{-1}$
    \cite{Abulencia:2006ze} & $17.77\ \mathrm{ps}^{-1} \pm 30\%$ \\
    \multirow{2}*{$\phi_{B_s}$} & $-0.57^{+0.24}_{-0.30}{}^{+0.07}_{-0.02}$
    \cite{Abazov:2008fj}
    & $[-1.20, 0.06]$ \\ 
    & $-0.76^{+0.37}_{-0.33}$, $-2.37^{+0.33}_{-0.37}$ \cite{Barberio:2008fa}
    & $[-1.26, -0.13] \cup [-3.00, -1.88]$ \\
    $B(\bsg)$        & $(352 \pm 23 \pm 9) \times 10^{-6}$
    \cite{Barberio:2008fa} & $2\ \sigma$ \\ 
    $\SphiK$     & $0.39 \pm 0.17$ \cite{Barberio:2008fa} & $2\ \sigma$ \\
    $|\epsilon_K|$     & $(2.232 \pm 0.007)\times 10^{-3}$ \cite{PDG} &
    $|\epsilon_K^\mathrm{SUSY}| < |\epsilon_K^\mathrm{exp}|$ \\
    $\epsilon'/\epsilon_K$ & $ (1.66 \pm 0.26)\times 10^{-3}$ \cite{PDG} &
    $|(\epsilon'/\epsilon_K)^\mathrm{SUSY}| <
    |(\epsilon'/\epsilon_K)^\mathrm{exp}|$ \\
    $|d_n|$ & $< 6.3\times 10^{-26}\ e\,\mathrm{cm}$ \cite{Harris:1999jx} \\
    \hline
  \end{tabular}
  \caption{Constraints from the quark sector on sfermion mixing.
    An empty third column means that the second column is used as is.}
  \label{tab:hconstraints}
}

The mass splittings of $B^0$ and $B_s$ mesons have been measured
with high precision.
The error of $\Delta M_{B_d}$ is 0.8\% and
that of $\Delta M_{B_s}$ is 0.7\%.
However, their theoretical prediction from
short-distance physics is not so precise.
The main obstacle stems from $f_{B_d}^2 B_{B_d}$
($f_{B_s}^2 B_{B_s}$) which enters the hadronic matrix element of
\bdbdbar\ (\bsbsbar) mixing,
parametrizing long-distance QCD effects.
The present uncertainty in lattice QCD calculation is around 30\%
(see e.g.\ \cite{Lenz:2006hd} and references therein).
A popular way to avoid this large uncertainty is to take
the ratio $\Delta M_{B_s} / \Delta M_{B_d}$
since the error in
$(f_{B_s}^2 B_{B_s}) / (f_{B_d}^2 B_{B_d})$ is much smaller.
Still, the SM prediction of the mass difference ratio has
an uncertainty of about 40\% due to the errors in the CKM matrix elements
\cite{Lenz:2006hd}.
As a comprehensive way to embrace the above uncertainties,
we require that each of computed
$\Delta M_{B_d}$ and $\Delta M_{B_s}$ falls within 30\% of
its central value, fixing $f_{B_d}^2 B_{B_d}$ and $f_{B_s}^2 B_{B_s}$.
In spite of the seemingly loose conditions,
we will find that
these requirements play impressive roles, given higher soft scalar mass.
The uncertainty decreases with the progress of lattice QCD,
and is estimated to be reducible down to 8--10\% with 6--60 tera flops year
of computing power \cite{latticeQCDestimation}.
The improved constraint from this smaller error is considered as well.

Although $\sin 2\beta$ does not suffer from
uncertainty in the $\Delta B = 2$ matrix element,
its SM prediction depends on
$V_{ub}$, which has a sizable error.
When we require $\sin 2\beta$ to be within the $2\ \sigma$ range
of its experimental value,
we allow for a $2\ \sigma$ variation in
$|V_{ub}| = (4.31 \pm 0.30) \times 10^{-3}$ \cite{PDG} as well.
As with the magnitude of mixing, we estimate effects of
an improved measurement of $\sin 2\beta$ at a super $B$ factory,
under the assumption that
it will converge to its SM value.
We use 2\% as a projected error of $|V_{ub}|$
and 0.005 as a stadard deviation of $\sin 2\beta$ \cite{Bona:2007qt}.
Given these smaller errors, the central values of $|V_{ub}|$ and $\sin 2\beta$,
if they remain as they are now, become
inconsistent with each other, reflecting the present tension between them.
Under this condition, the future $\sin 2\beta$ measurement would
appear to exclude the SM, and therefore
it would be hard to evaluate the influence of its improved precision.
However, there is a claim that the tension can be reconciled within
the SM \cite{Neubert:2008cp}.
We do not regard this as a signal of new physics,
and assume that $|V_{ub}|$ will decrease so that it becomes
compatible with the present $\sin 2\beta$.

As for $\phi_{B_s}$, the phase of \bsbsbar\ mixing,
we try two distinct ways of imposing the constraint:
(a) using the latest data from D\O\ at 90\% confidence level (CL)\@;
(b) employing the 90\% CL range
recently reported by the Heavy Flavor Averaging Group (HFAG)\@.
Regarding option (b), we choose the one obtained with constraints
from flavor-specific $B_s$ lifetime and $B_s$ semileptonic asymmetry.
(What is denoted by $\phi_{B_s}$ in this work is
$\phi_s^{J/\psi\phi}$ in the notation of HFAG\@.)
We present both of these cases
as they lead to very different impressions of the results---%
the D\O\ range includes the SM prediction of $\phi_{B_s}$ and hence
it still works as a bound on the room for new physics, while
the HFAG range lies outside the SM value, thereby indicating
the size of extra contribution required to account for the discrepancy
\cite{Parry:2007fe,lfvutfit}.
In order to compare the power of $\phi_{B_s}$ measurements at LHCb
with that of LFV,
we suppose that the future central value of $\phi_{B_s}$ is given by the SM,
despite the current hint of new physics at the level around $2\ \sigma$.
We assume that the error of $\phi_{B_s}$ will be 0.009
at $10\ \mathrm{fb}^{-1}$ \cite{phisLHCb}.

Measurement of the inclusive branching fraction $B(\bdg)$
had not been available until its preliminary result
was recently reported from BaBar \cite{LP07nakao}.
The precision is still low.
Considering the experimental and theoretical uncertainties,
we take modest upper and lower bounds guesstimated from
the exclusive branching fraction
$B(B \rightarrow \rho/\omega\,\gamma) = (1.18 \pm 0.17)\times 10^{-6}$
\cite{Barberio:2008fa}.
Unlike $\bdg$, the branching ratio of $\bsg$
has been measured with a high precision.
We impose a $2\ \sigma$ constraint on it.

We use QCD factorization \cite{BBNS} to evaluate $\SphiK$,
the sine term coefficient in the time-dependent $CP$ asymmetry of
$B \rightarrow \phi K$ \cite{SphiKBBNS,Kane:2002sp}.
This approach has a source of hadronic uncertainty stemming from
regularizing a divergent integral in the annihilation contribution.
We follow the original prescription in Ref.~\cite{BBNS},
i.e.\ we replace $\int_0^1 dy/y$ by
$X_A = (1 + \varrho\, e^{i \varphi}) \ln (m_B / \Lambda_h)$,
with $\Lambda_h = 500\MeV$, $0 \le \varrho \le 1$, and $0 \le \varphi < 2\pi$
\cite{Kane:2002sp}.
We regard $\SphiK$ as consistent with the data
if it is less than $2\ \sigma$
away from the central value for any $\varrho$ and $\varphi$.

We also incorporate the $CP$ violation parameters
$\epsilon_K$ and $\epsilon'/\epsilon_K$ in $K_L \rightarrow \pi\pi$
in the list.
Although we do not explicitly scan over a 1--2 mixing,
kaon physics can be influenced by double or higher order insertions.
Imposing those constraints,
we require that new physics contribution to each does not exceed
the measured value in size.
Concerning $\epsilon_K$, its prediction from squark mixing
can be made with an uncertainty
much smaller than $|\epsilon_K^\mathrm{exp}|$.
However, we are assuming that
there may be an arbitrary 1--2 squark mixing,
although we do not make a plot for it.
This is why we are using a rather conservative bound.

Finally, we examine $d_n$ the neutron electric dipole moment (EDM).
Recently it has been pointed out that
this observable can be greatly influenced if
there are both $LL$ and $RR$ down-type squark mixings at the same time
\cite{neutronEDM}.
In order to evaluate $d_n$,
we add contributions through the down quark EDM,
down quark chromoelectric dipole moment (CEDM), and
strange quark CEDM.

\TABLE[t]{%
  \renewcommand{\arraystretch}{1.1}
  \begin{tabular}{ccc}
    \hline
    Mode      & Present bound & Future bound      \\
    $B(\meg)$ & $1.2\times 10^{-11}$ 
    \cite{Brooks:1999pu} & $10^{-13}$
    \cite{MEG} \\
    $B(\teg)$ & $1.1\times 10^{-7}$ \cite{Aubert:2005wa} &
    $10^{-8}$ \cite{Hashimoto:2004sm}, $2\times 10^{-9}$ \cite{Bona:2007qt} \\
    $B(\tmg)$ & $4.5\times 10^{-8}$ \cite{Hayasaka:2007vc} &
    $10^{-8}$ \cite{Hashimoto:2004sm}, $2\times 10^{-9}$ \cite{Bona:2007qt} \\
    \hline
  \end{tabular}
  \caption{Constraints from radiative LFV decay modes.}
  \label{tab:lconstraints}
}%
\TABLE{%
  \renewcommand{\arraystretch}{1.1}
  \begin{tabular}{lc}
    \hline
    Observable       & Measured value
    \\
    $\SKstargam$ & $-0.19 \pm 0.23$ \cite{Barberio:2008fa}
    \\
    $\Srhogam$   & $-0.83 \pm 0.65 \pm 0.18$ \cite{Belle:2007jf}
    \\
    $\SBsKsgam$
    \\
    $\ACPbsgam$ & $0.004 \pm 0.037$ \cite{Barberio:2008fa}
    \\
    $\ACPbdgam$
    \\
    $\ACPbsdgam$
    \\
    \hline
  \end{tabular}
  \caption{Monitored observables.
    Precisions attainable at a super $B$ factory are summarized
  in Table~\ref{tab:monitorresult}.}
  \label{tab:monitored}
}%
We use the constraints from the lepton sector listed in
Table~\ref{tab:lconstraints}.
The second column shows the present 90\% CL
upper bound on each mode.
The third column is the prospective upper bound from future experiments.
The new limit on $\meg$ is the goal of MEG at 90\% CL\@.
Also, higher sensitivity to $\teg$ and $\tmg$
is anticipated from a super $B$ factory.
In Section~\ref{sec:regions}, we choose to use $10^{-8}$ as the future
limit on $B(\teg)$ and $B(\tmg)$, between the two numbers in each row
of the table.
If one wants to use $2\times 10^{-9}$ instead,
the result can be obtained easily:
multiply the upper bound on a given mass insertion from $\teg$ or $\tmg$,
by $1/\sqrt{5}$.

Imposing the conditions enumerated above,
we estimate possible deviations in additional observables
of interest, shown in Table~\ref{tab:monitored}.
The first three measure
time-dependent $CP$ asymmetries in radiative $B$ decays.
The definition of $\SKstargam$
is given by  \cite{Atwood:1997zr},
\begin{equation}
\begin{aligned}
  \mathcal{A}_{K^* \gamma} (t) \equiv & \,
  \frac{\Gamma (\overline{B_d} (t) \rightarrow K^* \gamma) -
        \Gamma (          B_d  (t) \rightarrow K^* \gamma)}%
       {\Gamma (\overline{B_d} (t) \rightarrow K^* \gamma) +
        \Gamma (          B_d  (t) \rightarrow K^* \gamma)}
      \\
      = & \ACPKstargam \cos (\Delta M_{B_d} t)
        + \SKstargam \sin (\Delta M_{B_d} t) .
\end{aligned}
\end{equation}
\noindent Note that the time-dependent $CP$ asymmetry in
$B_d \rightarrow K^* \gamma$ in our convention has the sign opposite to
that in the above reference.
This observable is sensitive to a new $CP$ violating phase
in the right-handed $b \rightarrow s$ transition, such as coming from
$\ded{23}{RR}$.
We define $\Srhogam$ in a parallel way by replacing $K^*$ with $\rho$
in the expression.
This can serve as a $b \rightarrow d$ analog of $\SKstargam$,
affected by $\ded{13}{RR}$.
One might as well use $\SBsKsgam$ to investigate $\ded{13}{RR}$,
and we record its variation.
The rest three are direct $CP$ asymmetries in radiative $B$ decays,
$B \rightarrow X_s \gamma$, $B \rightarrow X_d \gamma$, and
$B \rightarrow X_{s+d} \gamma$,
whose definitions can be figured out by setting $t = 0$
in the above equation.
They are complementary to the preceding observables
in the sense that they can probe left-handed $CP$ violating new physics
such as $\ded{23}{LL}$ and $\ded{13}{LL}$.
We quote the measured value of each observable if available.

\section{Results}
\label{sec:results}

\subsection{Viable region of each mass insertion}
\label{sec:regions}

As a preparation for reading plots of the GUT scale mass insertions,
we sketch the process amplitudes in terms of these variables.
This will help us understand how a figure changes
as a parameter is modified.
Keeping only factors of interest,
the LFV decay amplitudes can be roughly put in the form,
\begin{subequations}
  \label{eq:ALFV}
\begin{align}
A(\tmg) &\propto
\frac{\mu \tb\cdot\del{23}{LL}(\MGUT)}{m_S^2}  ,
\\
  \label{eq:Ameg}
A(\meg) &\propto
\frac{m_\tau\mu\tb}{m_0^2} \times
\frac{\del{13}{RR}(\MGUT)\cdot\del{32}{LL}(\MGUT)}{m_S^2}
,
\end{align}
\end{subequations}
where $m_S$ is the typical mass of a slepton, chargino, or neutralino
in the loop.
The first factor in the second line is
$\del{33}{RL} \equiv m_\tau (A - \mu \tb) / \msl^2$
rewritten with~\eqref{eq:sleptondiagonalrunning}.
We used~\eqref{eq:delLLRRRG} to replace the other mass insertions
by the GUT scale quantities.
As to hadronic observables, let us pick up $\Delta M_{B_s}$
as an example; other constraints can be understood in a similar fashion.
The \bsbsbar\ transition amplitude depends on
$\ded{23}{AA}\ded{23}{BB} / m_S^2$ with $A, B = L, R$.
For instance, we can use~\eqref{eq:squarkMIdilution}
to recast one of these combinations at $\MSUSY$ as
\begin{equation}
  \label{eq:d23LLd23RR}
  \left. \frac{\ded{23}{LL}\ded{23}{RR}}{m_S^2} \right|_{\MSUSY}
  \approx
  \frac{[\ded{23}{LL}(\MGUT)+q_{ij}]\cdot\ded{23}{RR}(\MGUT)}%
  {(1 + 6 x)^2 m_S^2}
  ,
\end{equation}
where $m_S$ is the typical mass of a squark or gluino in the loop.
Note that $q_{ij}$ from~\eqref{eq:qij} is nearly independent of $m_0^2$.
Therefore, as we vary $m_0$ and $M_{1/2}$,
the scaling property of~\eqref{eq:d23LLd23RR} is determined by
its denominator.
The other two combinations with $A = B = L, R$ scale in the same way.

\begin{figure}
  \centering
\subfigure[$m_0 = 220 \GeV,\ M_{1/2} = 180 \GeV,\ \tb = 5$]{\incgr[height=62mm]{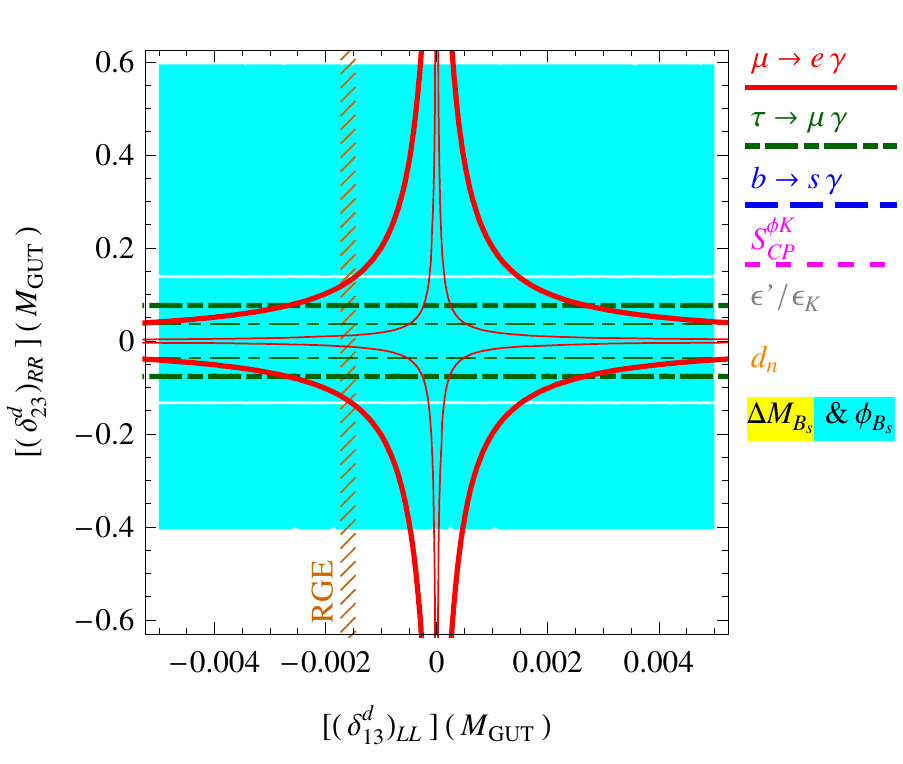}}
\subfigure[$m_0 = 600 \GeV,\ M_{1/2} = 180 \GeV,\ \tb = 5$]{\incgr[height=62mm]{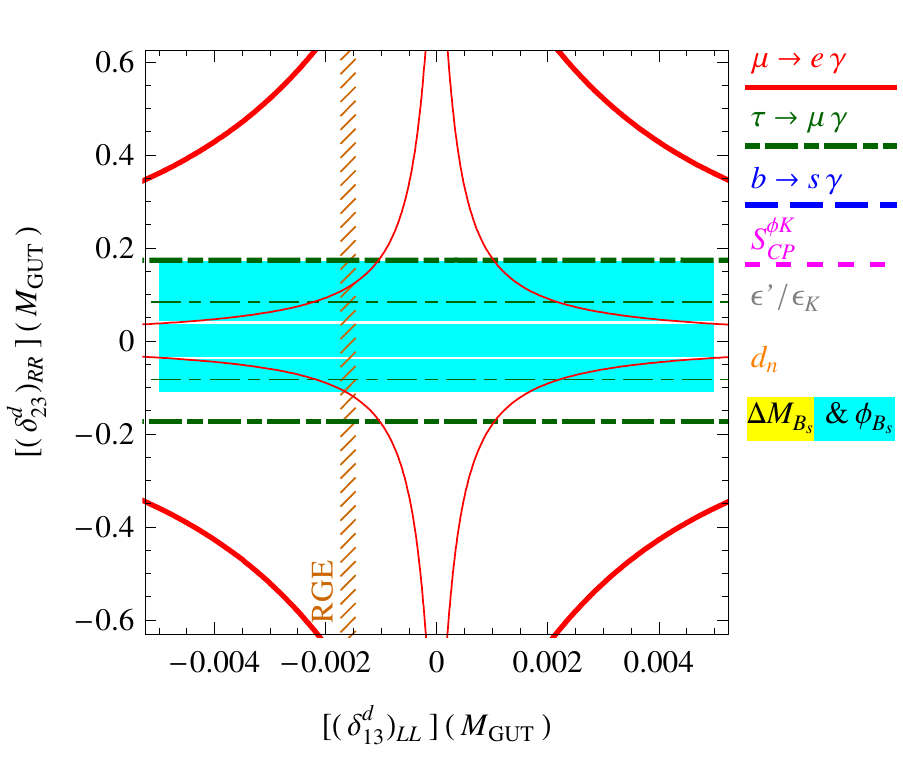}}
\\
\subfigure[$m_0 = 220 \GeV,\ M_{1/2} = 180 \GeV,\ \tb = 5$]{\incgr[height=63mm]{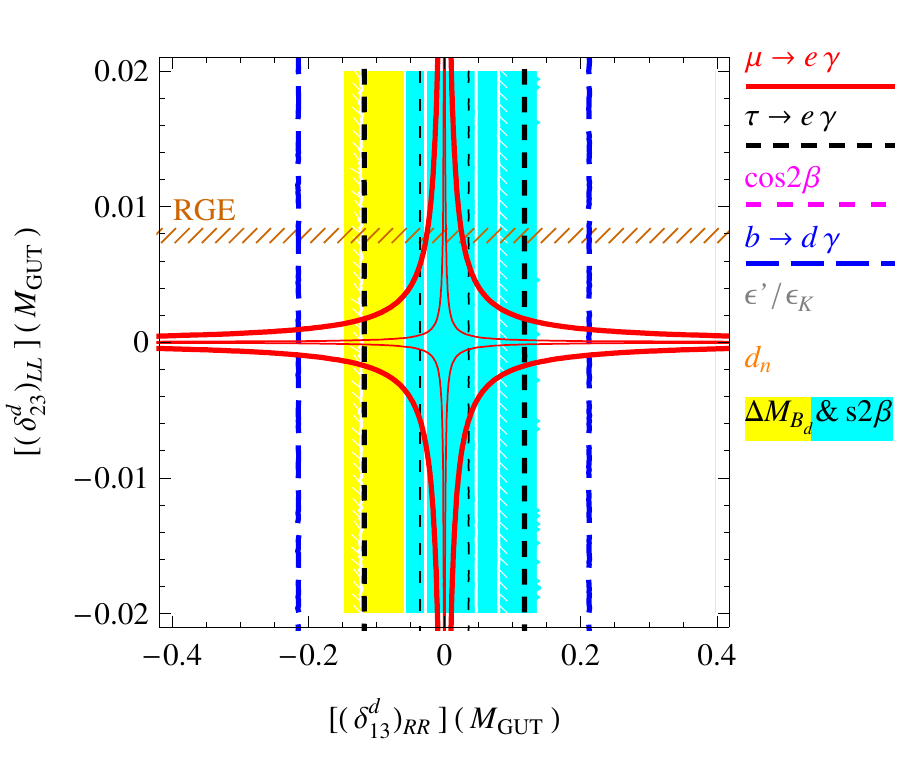}}
\subfigure[$m_0 = 600 \GeV,\ M_{1/2} = 180 \GeV,\ \tb = 5$]{\incgr[height=63mm]{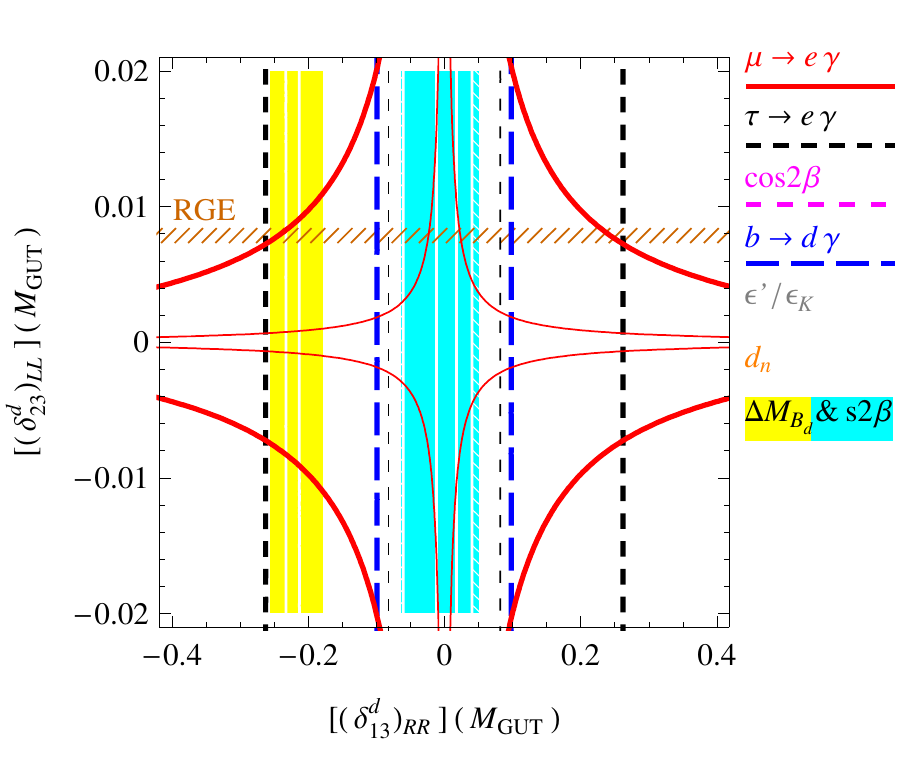}}
  \caption{Constraints on [(a), (b)] the $(\ded{13}{LL}, \ded{23}{RR})$
    and [(c), (d)] the $(\ded{13}{RR}, \ded{23}{LL})$ planes.
    For each LFV process,
    the thick curve is the present upper bound and
    the thin curve is the prospective future bound.
    A light gray (yellow) region is allowed by [(a), (b)] $\Delta M_{B_s}$ 
    or [(c), (d)] $\Delta M_{B_d}$,
    given 30\% uncertainty in the $\Delta B = 2$ matrix element,
    and a gray (cyan) region is further consistent with
    [(a), (b)] $\phi_{B_s}$ from D\O\ or [(c), (d)] $\sin 2 \beta$.
    The white lines around the center mark a possible improved
    constraint from [(a), (b)] $\Delta M_{B_s}$ or [(c), (d)] $\Delta M_{B_d}$,
    with 8\% hadronic uncertainty.
    The white curves with short thin lines attached to them
    display a measurement of [(a), (b)] $\phi_{B_s}$ at LHCb or
    [(c), (d)] $\sin 2 \beta$ at a super $B$ factory.
    Those short lines indicate the excluded side.
    Contributions from RG evolution to [(a), (b)] $\ded{13}{LL}$ and
    [(c), (d)] $\ded{23}{LL}$ are indicated by the vertical and the horizontal
    hatched strips, respectively.
    Their widths do not have any meaning.}
  \label{fig:twoparameterscan}
\end{figure}

With these ingredients at hand, we begin to interpret the results.
First of all, let us look at the result of a two-parameter scan
to get a taste of two mass insertions.
In each of the four plots shown in Figs.~\ref{fig:twoparameterscan},
there are 1--3 and 2--3 mixings with different chiralities.
The gaugino mass $M_{1/2}$ at the GUT scale
is chosen in such a way that
the gluino mass becomes 500 GeV at the weak scale.
The scalar mass $m_0$ at $\MGUT$ is set to two different values
that can elucidate complementarity of the quark and the lepton
sector processes.
In the left column, $m_0$ is taken to be $220\GeV$, so that
the first and the second family right-handed down-type squarks
have the same mass as the gluino at the weak scale.
(The third family is slightly lighter.)
This is a benchmark case often encountered in the literature
on supersymmetric flavor violation.
In the right column, we change $m_0$ to $600\GeV$.
If one fixes the $\delta$ parameters at $\MGUT$,
this $m_0$ maximizes gluino loop contribution to $B$-meson mixing
for the gaugino mass chosen here.
We elaborate on this point later.
On the plots, each mass insertion parameter is treated as a real number.

In each of Figs.~\ref{fig:twoparameterscan}~(a) and (b), the two axes are
$\ded{13}{LL} = \del{13}{RR}^*$ and $\ded{23}{RR} = \del{23}{LL}^*$
at the GUT scale.
Here we restrict the horizontal axis to a range much narrower than
the vertical axis since we are especially interested in the effect
of RG contribution, but otherwise the $LL$ mixing can be arbitrary.
One can find that $\ded{23}{RR}$
is constrained by $\tmg$ and $\Delta M_{B_s}$.
Interestingly, the two plots show
different relative significance of these two constraints.
In Fig.~(a), $\tmg$ is stronger than $\Delta M_{B_s}$, i.e.\
a large portion of the region allowed by the latter is excluded by
the former.
In Fig.~(b), the order of importance appears to be reversed.
Although it is early to draw a conclusion
since these plots restrict the mass insertions to be real,
it is obvious that $\Delta M_{B_s}$ gets tighter while
$\tmg$ becomes looser if $m_0$ is changed from $220\GeV$ to $600\GeV$.
As $m_0$ increases,
the LFV constraints get relaxed because $m_S$ gets bigger
in~\eqref{eq:ALFV}.
In fact, $\mu$ in a numerator grows as well,
but the growth of $m_S^2$ in the denominator wins.
In spite of heavier sparticles, hadronic constraints get
relatively more stringent.
To understand why, pay attention to~\eqref{eq:d23LLd23RR}.
Since we have fixed the gluino mass while raising $m_0$,
a squark is heavier than a gluino, and
we should substitute~\eqref{eq:squarkdiagonalrunning} for $m_S^2$.
Then, the gluino loop contribution to
\bsbsbar\ mixing scales like $x / (1 + 6 x)^3$.
This factor increases as we decrease $x$
inversely proportional to $m_0^2$,
unless $x$ is smaller than $1/12$, the maximum point.
For $x \lesssim 1/12$, the squarks are so heavy
that they begin to decouple from low-energy processes
as in split supersymmetry.
Note that $x = 0.67$ on the left plot and $x = 0.09$ on the right.
This explains the narrower $\Delta M_{B_s}$ band on the right plot.\footnote{%
  A similar discussion is given in Ref.~\cite{Dutta:2008xg}
  in a different context, in a more qualitative way.
  In their scenario, the 2--3 squark mixing arises from
  large neutrino Yukawa couplings.
  They state that squark loop effects on $B_s$ mixing can be more significant
  for higher $m_0$.
  However, they do not mention at what point of $m_0$
  this trend stops and squark loops begin to decouple.}
Other quark sector processes are enhanced in a similar way,
as will be shown later.
As can be expected from Fig.~\ref{fig:megtriple}, $\meg$
restricts the product $\ded{13}{LL} \ded{23}{RR}$, resulting in
the hyperbolas on the plane.
Therefore, the $\meg$ limit on $\ded{23}{RR}$ varies
depending on the size of $\ded{13}{LL}$.
If $\ded{13}{LL} = 0$, $\ded{23}{RR}$ is free,
but the restraint grows severer as $|\ded{13}{LL}|$ increases.
A special case with RG-induced $\ded{13}{LL}$,
marked by the vertical hatched strip, will be detailed shortly.
With increasing $m_0$,
$\meg$ is doubly suppressed by $m_0^2\,m_S^2$ in~\eqref{eq:Ameg}.
This expands the area within the hyperbola in the plots.
The width of the $\Delta M_{B_s}$ band is mainly due to
the current uncertainty
in the \bsbsbar\ mixing matrix element around 30\%.
The projected bound with 8\% uncertainty is depicted by
the two white lines around $\ded{23}{RR} = 0$.
In Fig.~(a), it is not yet as tight as the present $\tmg$ constraint
which will be even tighter in the future.
However, in Fig.~(b), the reduced hadronic uncertainty makes
the $\Delta M_{B_s}$ bound more restrictive than $\tmg$ at a super $B$ factory.
Nevertheless, it should be kept in mind that $\tmg$ becomes more sensitive
as $\tb$ grows while \bsbsbar\ mixing does not.

We do the same exercise with a different mixture of mass insertions,
$\ded{13}{RR} = \del{13}{LL}^*$ and $\ded{23}{LL} = \del{23}{RR}^*$,
to get Figs.~\ref{fig:twoparameterscan}~(c) and (d).
The vertical range is set around the magnitude generated by RG running.
Here $\ded{13}{RR}$ is bounded by $\Delta M_{B_d}$, $\sin 2\beta$,
$\teg$, and $\bdg$.
In Fig.~(c), the current limits from $\Delta M_{B_d}$ and $\teg$
are comparable to each other, which are stronger than $\bdg$.
In Fig.~(d), $\Delta M_{B_d}$ with the aid of $\sin 2\beta$
leaves a band which is much narrower than that allowed by $\teg$.
Also, the $\bdg$ bound moves inside the $\teg$ bound.
These changes, as well as enhancement of
the other quark sector processes,
can be understood in the same way as the difference between Figs.~(a) and (b).
The inner two white lines around $\ded{13}{RR} = 0$ indicate
the limit from $\Delta M_{B_d}$ with 8\% uncertainty
in the \bdbdbar\ mixing matrix element,
and the outer two white lines with short thin lines attached to them
arise from $\sin 2\beta$ at a super $B$ factory.
The projected limits from $\Delta M_{B_d}$ and $\teg$
are expected to maintain the present tendency of relative strengths.
That is, they are comparable to each other in Fig.~(c)
and the former is stronger than the latter in Fig.~(d).
Again, the hyperbolas come from $\meg$.
Boundaries by the neutron EDM appear in Fig.~(d) even though
all the mixings are real.
Extra contribution to $d_n$ arises from the combination
$\ded{13}{LL} \ded{13}{RR}^*$ of which the $LL$ insertion picks up
a non-vanishing phase from the CKM matrix, running from the GUT scale
down to the weak scale.
A specific case with RG-induced $\ded{23}{LL}$,
indicated by the horizontal hatched strip, will be discussed later.

Having grasped a picture of how different constraints
act on two mass insertions,
let us examine cases where one of the insertions originates from
RG running from the reduced Planck scale to the GUT scale.

\begin{figure}
  \centering
\subfigure[$m_0 = 220 \GeV,\ M_{1/2} = 180 \GeV,\ \tb = 5$]{\incgr[height=63mm]{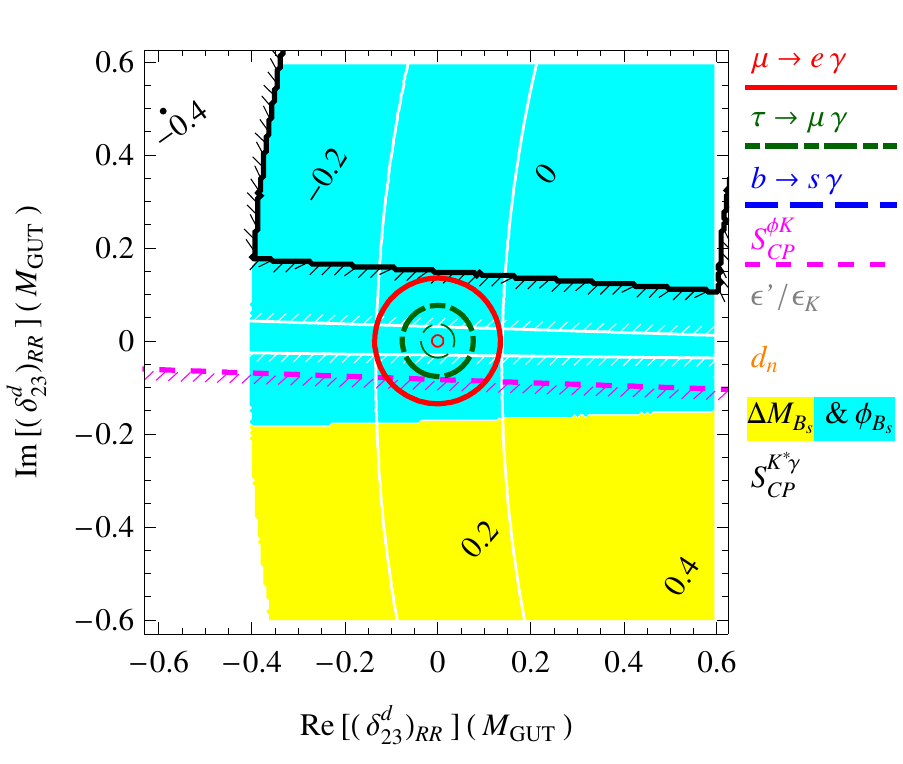}}
\subfigure[$m_0 = 220 \GeV,\ M_{1/2} = 180 \GeV,\ \tb = 10$]{\incgr[height=63mm]{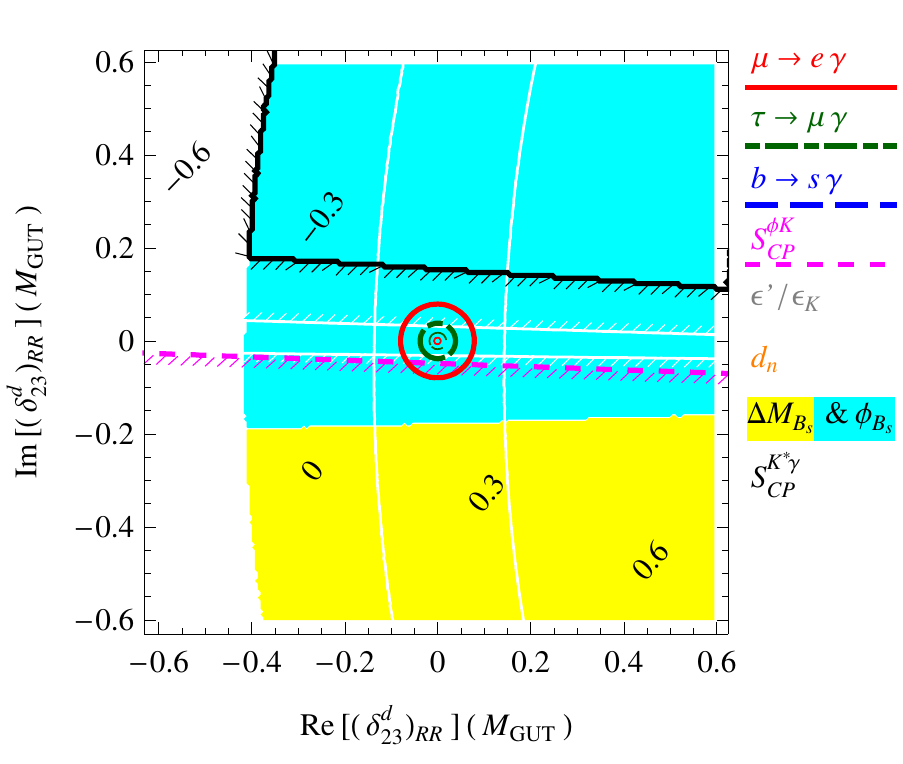}}
\\
\subfigure[$m_0 = 600 \GeV,\ M_{1/2} = 180 \GeV,\ \tb = 5$]{\incgr[height=63mm]{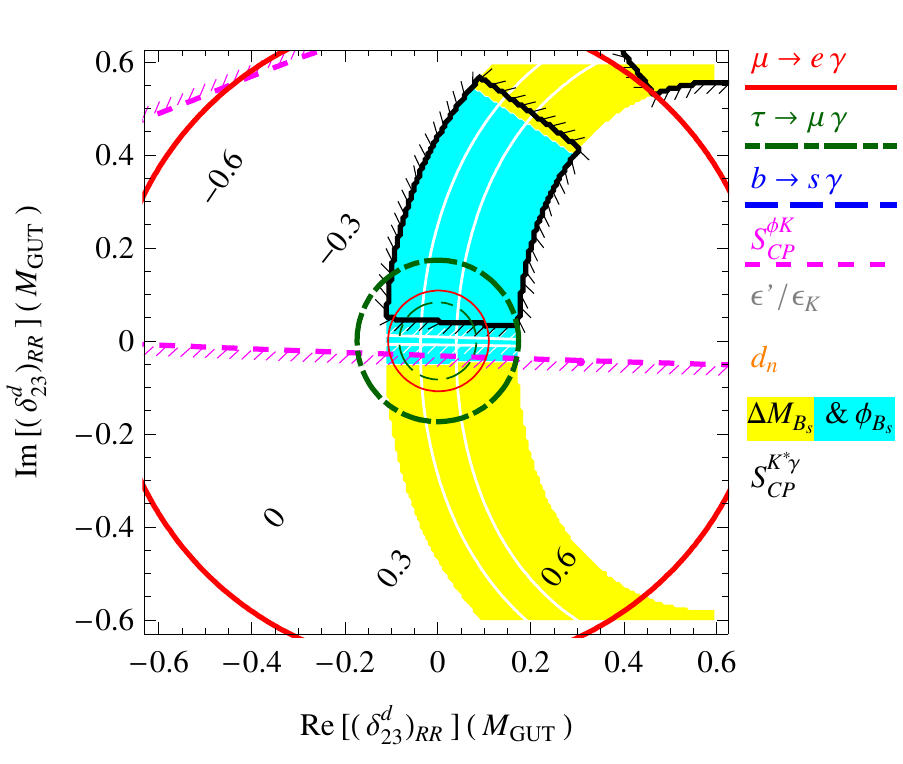}}
\subfigure[$m_0 = 600 \GeV,\ M_{1/2} = 180 \GeV,\ \tb = 10$]{\incgr[height=63mm]{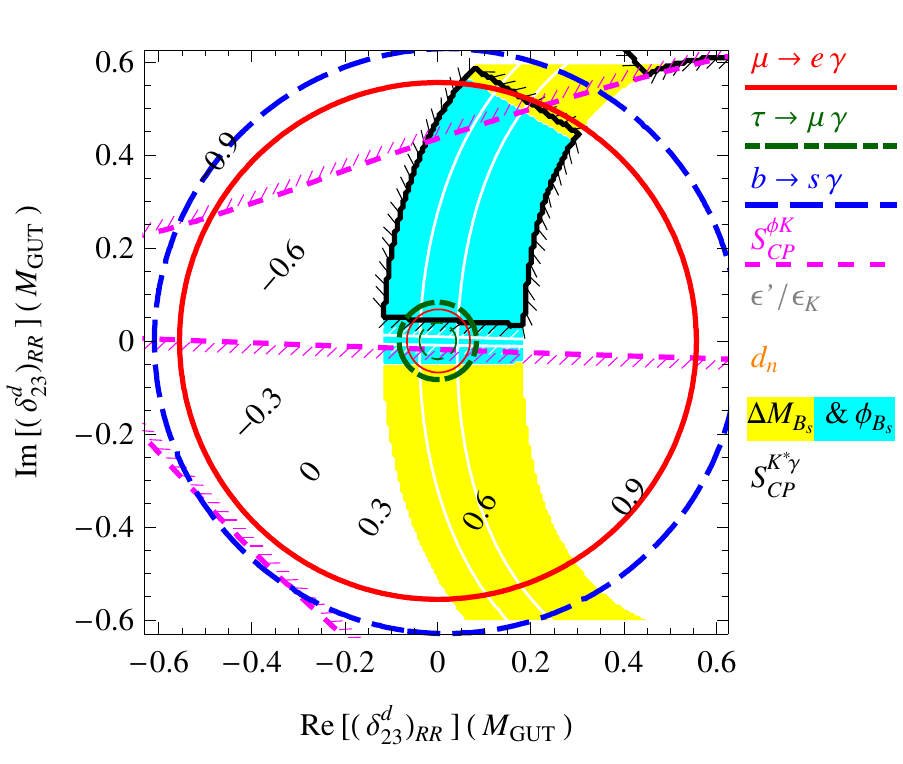}}
  \caption{Constraints on the complex plane of $\ded{23}{RR}$,
    with $\ded{ij}{LL}$ generated from
    RG running between the reduced Planck scale and the GUT scale.
    For each LFV process,
    the thick circle is the present upper bound and
    the thin circle is the prospective future bound.
    A light gray (yellow) region is allowed by $\Delta M_{B_s}$,
    given 30\% uncertainty in the $\Delta B = 2$ matrix element,
    and a gray (cyan) region is further consistent with $\phi_{B_s}$
    from D\O.
    A thick black curve shows $\phi_{B_s}$ from HFAG\@.
    The white curves running from top to bottom mark a possible improved
    constraint from $\Delta M_{B_s}$
    with 8\% hadronic uncertainty.
    The other white lines running from left to right
    display a measurement of $\phi_{B_s}$ at LHCb.
    Thin short lines attached to a curve
    indicate the excluded side.}
  \label{fig:23RR}
\end{figure}

\begin{figure}
  \centering
  \subfigure[$m_0 = 220 \GeV,\ M_{1/2} = 180 \GeV,\ \tb = 5$]{\incgr[height=63mm]{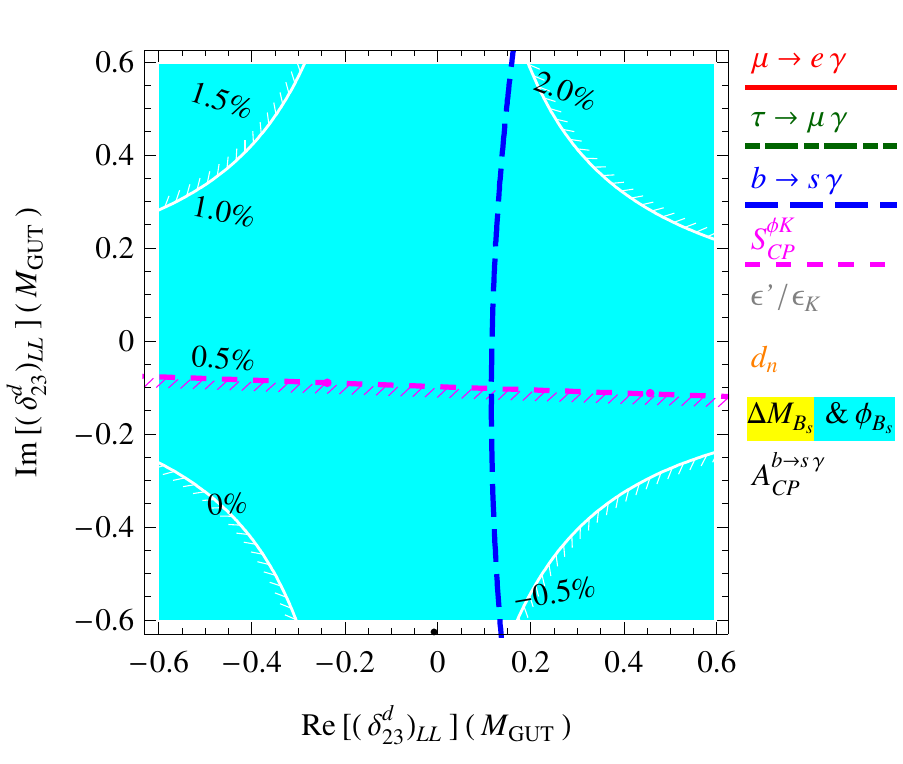}}
  \subfigure[$m_0 = 220 \GeV,\ M_{1/2} = 180 \GeV,\ \tb = 10$]{\incgr[height=63mm]{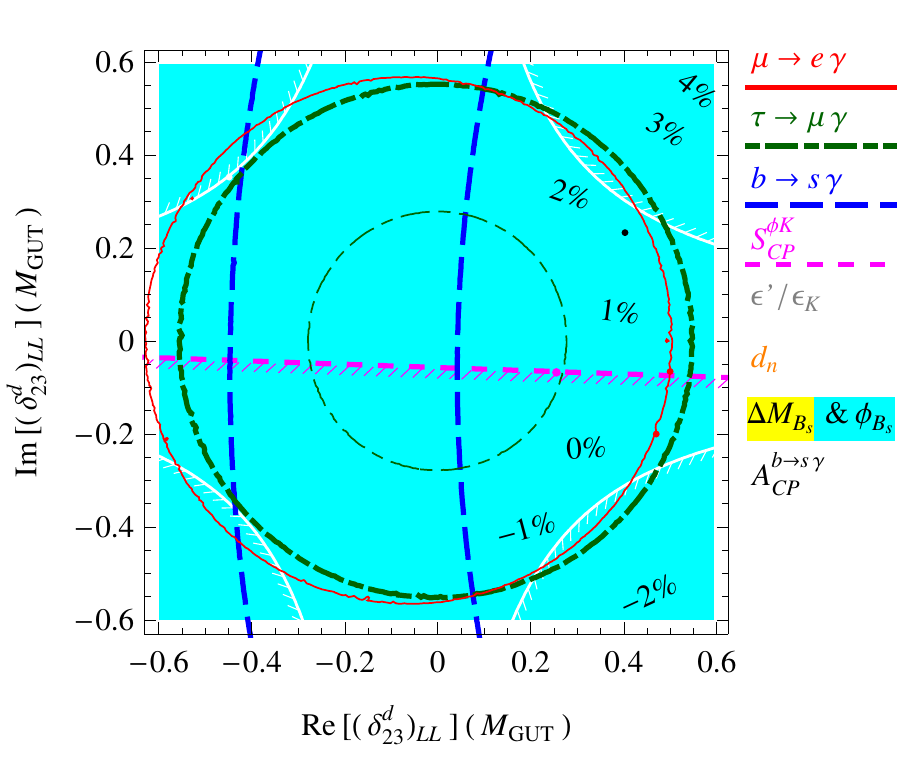}}
\\
  \subfigure[$m_0 = 600 \GeV,\ M_{1/2} = 180 \GeV,\ \tb = 5$]{\incgr[height=63mm]{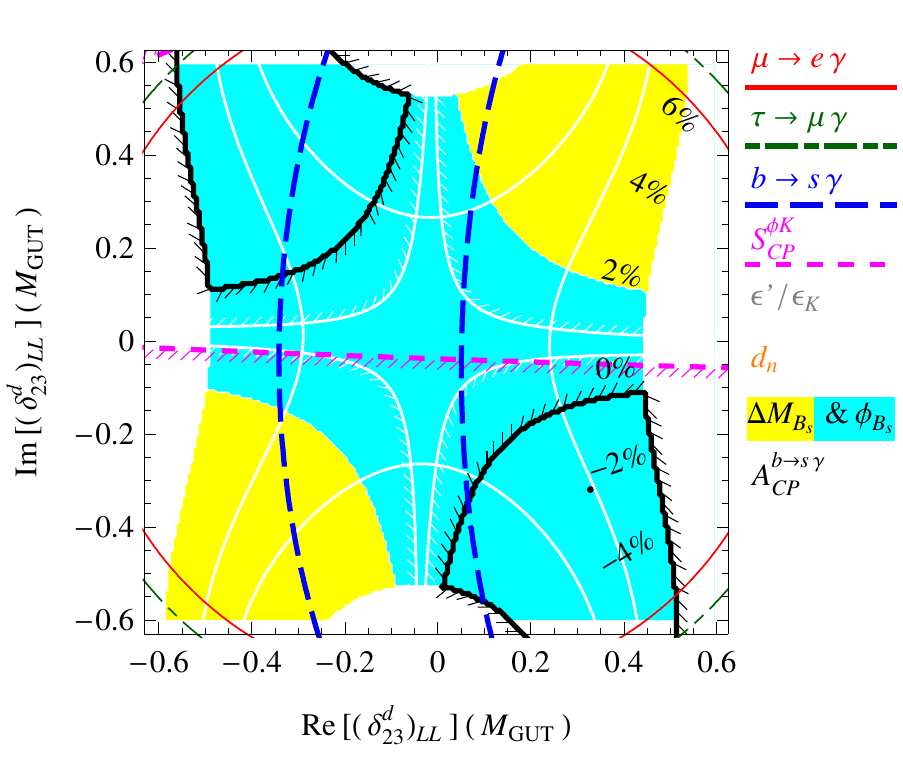}}
  \subfigure[$m_0 = 600 \GeV,\ M_{1/2} = 180 \GeV,\ \tb = 10$]{\incgr[height=63mm]{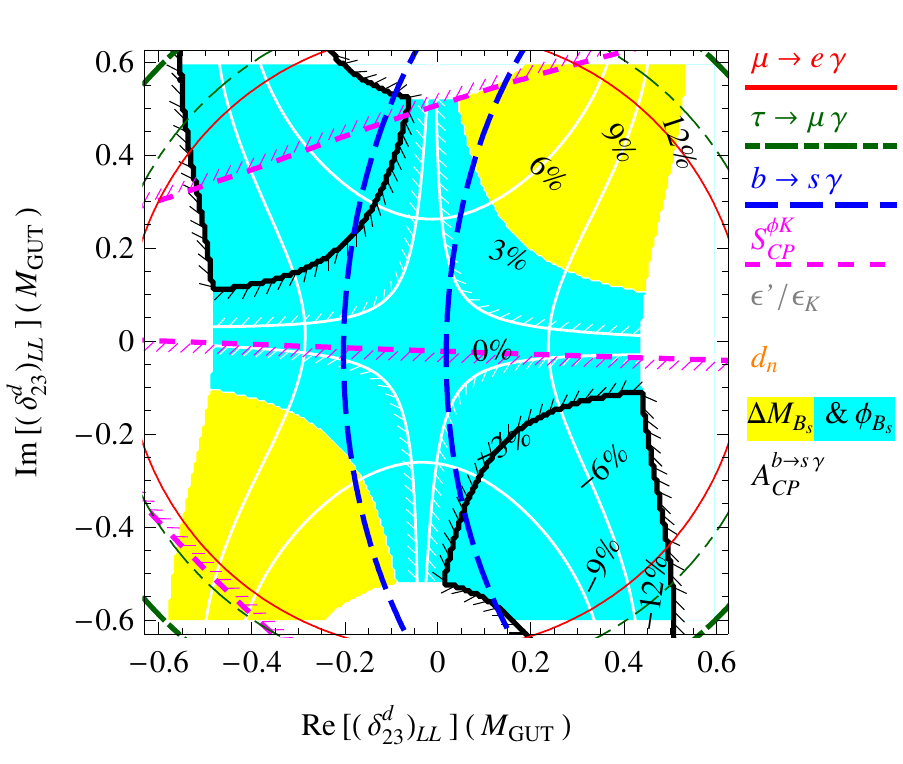}}
  \caption{Constraints on the complex plane of $\ded{23}{LL}$,
    with $\ded{12}{LL}$ and $\ded{13}{LL}$ generated from
    RG running between the reduced Planck scale and the GUT scale.
    For each LFV process,
    the thick circle is the present upper bound and
    the thin circle is the prospective future bound.
    A light gray (yellow) region is allowed by $\Delta M_{B_s}$,
    given 30\% uncertainty in the $\Delta B = 2$ matrix element,
    and a gray (cyan) region is further consistent with $\phi_{B_s}$
    from D\O.
    A thick black curve shows $\phi_{B_s}$ from HFAG\@.
    The white curves without short thin lines
    mark a possible improved
    constraint from $\Delta M_{B_d}$
    with 8\% hadronic uncertainty.
    The white curves with short thin lines attached to them
    display a measurement of $\phi_{B_s}$ at LHCb.
    Thin short lines attached to a curve
    indicate the excluded side.}
  \label{fig:23LL}
\end{figure}

We display in Figs.~\ref{fig:23RR}, complex versions of
Figs.~\ref{fig:twoparameterscan}~(a) and (b) with the $LL$ insertions fixed at
the numbers shown in Table~\ref{tab:insertions}.
They indicate regions of $\ded{23}{RR} = \del{23}{LL}^*$ (in)consistent with
observations.
This time, $\tb$ is varied as well as $m_0$.
Let us walk through them starting from Fig.~\ref{fig:23RR}~(a).
A light gray (yellow) area is allowed by $\Delta M_{B_s}$
but not by $\phi_{B_s}$ from D\O\@.
A gray (cyan) area is allowed by both.
However, most of it is ruled out by the LFV processes,
as we have already noticed in Fig.~\ref{fig:twoparameterscan}~(a).
One can guess that this should be the case even in the near future,
comparing the zone surrounded by the white curves and the thin circles
with their centers at the origin.
In particular, the $\meg$ data from the MEG experiment
should be able to kill all the parameter space except for the tiny disk
around the origin.
It deserves a remark that $\meg$ is
playing an important role here.
Being a 2--3 mixing, $\ded{23}{RR}$ is normally
associated with the $\tmg$ process.
For example, Ref.~\cite{Ciuchini:2007ha} discusses interplay between
leptonic and hadronic constraints in a similar context, but they use
only $\tau \rightarrow \mu$ transitions to restrict $\ded{23}{RR}$.
This difference arises from the strategy of
setting the mass insertion parameters.
Their default value of a mass insertion is zero,
while our default is the one which is minimally expected from
RG running.
Therefore, they do not find $\meg$ limiting a 2--3 mixing
as is obvious from Fig.~\ref{fig:twoparameterscan}~(a).
We believe that our choice of mass insertions is more reasonable
in a scenario where the soft terms are generated
around the Planck scale such as gravity mediation.
It may be argued that the RG-induced $\mathbf{10}$ scalar mixing
is not always guaranteed to be sizeable
since the cutoff scale can happen to be low close to the GUT scale.
This is true.
However, a low cutoff would threaten the validity of making a connection
between the quark and the lepton flavors in the first place.
Non-renormalizable operators shown in~(\ref{eq:wnr})
and even higher order terms,
generically, give $\order (1)$ contributions to the quark and the lepton
Yukawa couplings, thereby erasing any trace of their connection
in the flavor space as a single GUT multiplet.
It should be remembered that
an RG-induced $LL$ insertion is not only critical to $\meg$, but also
to \bsbsbar\ mixing.
Indeed, the presence of $\ded{23}{LL}$ is rendering
the $\Delta M_{B_s}$ constraint on $\ded{23}{RR}$ tighter
\cite{Ciuchini:2006dx,Endo:2006dm,Gabbiani:1988rb}.
If it were not for $\ded{23}{LL}$, the gray region would look like
the one in Fig.~\ref{fig:23LL}~(a),
where the contribution to $\Delta M_{B_s}$ from $\ded{23}{LL}$
is not enhanced by $\ded{23}{RR}$.
Another noticeable point is that the information on $\phi_{B_s}$ from LHCb
can play an important role in shaping the allowed region.
A particular pleasure with this constraint
that it does not suffer from the hadronic uncertainty that
plagues $\Delta M_{B_s}$.
Other observables of interest related to 2--3 mixing are
$\SphiK$ and $d_n$.
The area excluded by each of them is depicted.
Note that $d_n$ depends on the phase of $\ded{23}{LL}$ as well as
on its size.
Although the phase of the first term in~\eqref{eq:dedLLRG}
is fixed by the CKM matrix elements, the $\order(\xi)$
correction is unknown and may influence the phase of the entire insertion.
Varying the phase of $\ded{23}{LL}$ amounts to rotating the $d_n$ band
on the plot around the origin.
Finally, the dotted lines are contours of $\SKstargam$.
 From them, one can read off its largest possible deviation that is consistent
with the other experimental inputs.
Further information on this $CP$ asymmetry is collected
in Section~\ref{sec:otherobs}.

Now that we have recognized the general structure of a plot,
we try different values of parameters.
First, $\tb$ is doubled from 5 to 10 in Fig.~\ref{fig:23RR}~(b).
The $\Delta M_{B_s}$ belt does not change very much since
its dependence on $\tb$ is negligible.
Each LFV circle halves and becomes tighter.
This is evident from~\eqref{eq:ALFV}, where
each decay amplitude is proportional to $\tb$.
The gluino loop diagrams contributing to each of $d_n$ and
the $B \rightarrow \phi K$ decay are also proportional to $\tb$,
and the allowed region shrinks as $\tb$ increases.
Next, we change the scalar mass parameter.
In Fig.~\ref{fig:23RR}~(c), we raise $m_0$ to $600\GeV$,
a value optimized for $B$-meson mixings.
For the reason already explained,
the LFV constraints turn weaker while
the \bsbsbar\ mixing belt shrinks on the plot.
Other quark sector processes are boosted as well.
Because of this, the impressions of hadronic and leptonic
bounds undergo a sea change from Fig.~(a) to (c).
The current $\meg$ limit gets so much relaxed that
it does not exclude any region compatible with
$\Delta M_{B_s}$ and $\phi_{B_s}$.
One can also notice that $d_n$ has become much more powerful.
Its limit on the imaginary part of $\ded{23}{RR}$ is
stronger than any other bound on the plot.
Indeed, the combination of $d_n$ and \bsbsbar\ mixing leaves
nothing to do for $\tmg$.
In the future case, \bsbsbar\ mixing looks more restrictive than
$\tmg$ and $\meg$, particularly thanks to
improved precision of $\phi_{B_s}$ at LHCb.
This should be contrasted with the situation in Fig.~(a)
where the LFV constraints, both at present and in the future,
are stronger than the hadronic ones.
Lastly, we consider higher $m_0$ and $\tb$ in Fig.~(d).
Each observable changes according to its $\tb$ dependence
already mentioned.
For the first time, the $\bsg$ bound becomes visible.
Until now, its branching fraction has not been sufficiently disturbed by
the new physics contribution given by Fig.~\ref{fig:bsg}~(b).
One reason is that this diagram does not interfere with
the SM one since they lead to different photon helicities.
In Fig.~\ref{fig:23RR}~(d), however,
higher $m_0$ and $\tb$ cooperate to enhance
the supersymmetric amplitude.
Still, $\bsg$ is not very restrictive.
Its role should be more significant for $\tb$ much higher than 10.
The range of $\SKstargam$ predicted in each case is summarized
in Table~\ref{tab:monitorresult}.

A remark is in order regarding the recent reports on $\phi_{B_s}$
which reveal a small but interesting disparity between
the combined fit and the SM prediction \cite{Bona:2008jn,Barberio:2008fa}.
Once we take the 90\% CL range of $\phi_{B_s}$ from HFAG, instead of
that from D\O, the gray regions change to those surrounded
by the thick black curves.
Since the HFAG result demands new physics contribution to $\phi_{B_s}$,
the origin on the plane is positioned outside the thick black boundary.
On the other hand, the mass insertion can be compatible with the LFV data
only around the origin.
This conflict leads to a restriction in an attempt to understand
the new fit result of $\phi_{B_s}$ with an $RR$ mixing.
The trouble is more serious with lower $m_0$ and/or higher $\tb$.
As was explained above, lower $m_0$ enhances
the LFV branching ratios while
suppressing supersymmetric contributions to \bsbsbar\ mixing.
Also, higher $\tb$ gives rise to higher LFV rates.
Figs.~(a) and (b) tell us that
these cases are disfavored by LFV in combination with $\phi_{B_s}$.
The tension between LFV and $\phi_{B_s}$ is relaxed for higher $m_0$
used in the lower plots.
Indeed, one can find a small intersection of the $\phi_{B_s}$ area
and the $\tmg$ disk in each of Figs.~(c) and (d).
This region could be accessed by measurements of
$\tmg$ and $\meg$ in the near future.
However, the neutron EDM becomes a new obstacle to
the zone favored by $\phi_{B_s}$
as higher $m_0$ reinforces hadronic constraints.
The limit from $d_n$ grows more serious for higher $\tb$.
The problem can be eased by modifying the size and phase of
$\ded{23}{LL}$ at the GUT scale to rotate the $d_n$ band
as already mentioned.

\begin{figure}
  \centering
  \subfigure[$m_0 = 220 \GeV,\ M_{1/2} = 180 \GeV,\ \tb = 5$]{\incgr[height=63mm]{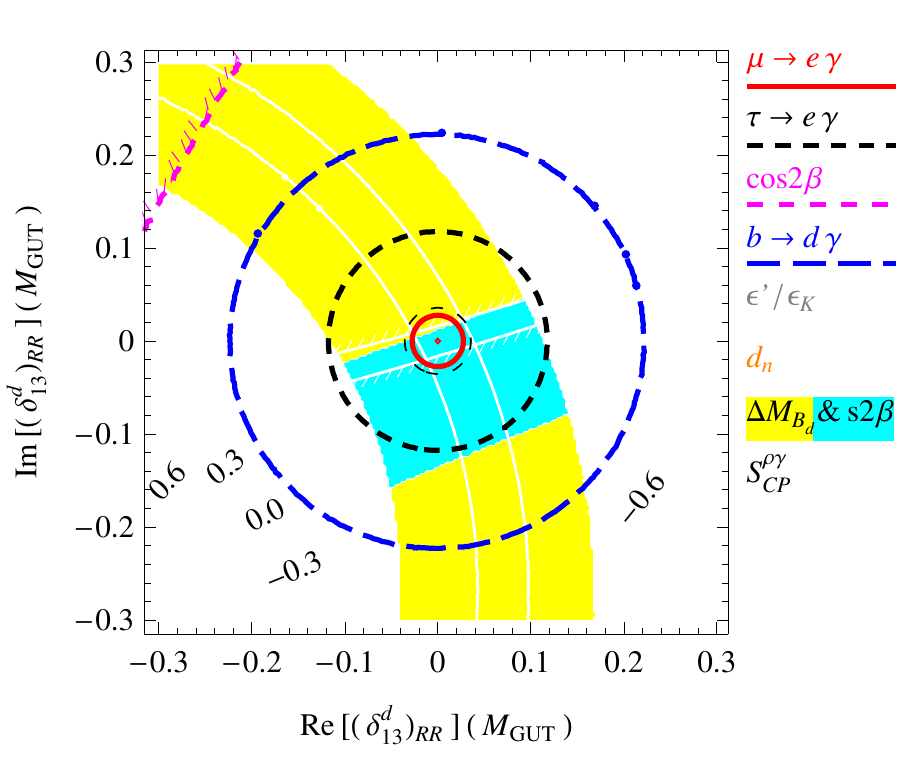}}
  \subfigure[$m_0 = 220 \GeV,\ M_{1/2} = 180 \GeV,\ \tb = 10$]{\incgr[height=63mm]{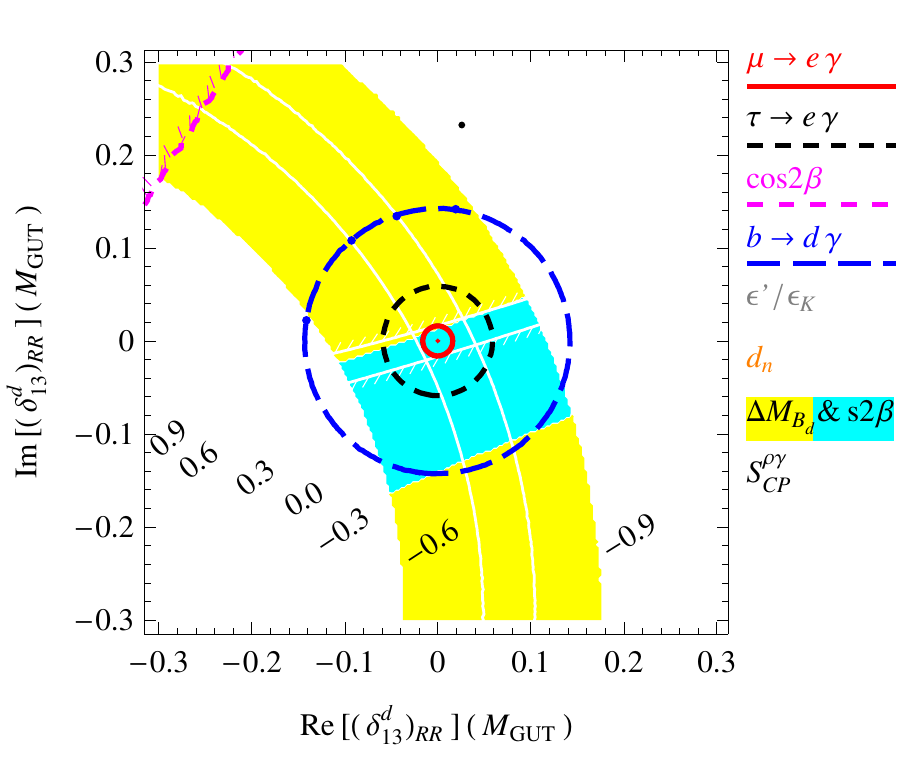}}
\\
  \subfigure[$m_0 = 600 \GeV,\ M_{1/2} = 180 \GeV,\ \tb = 5$]{\incgr[height=63mm]{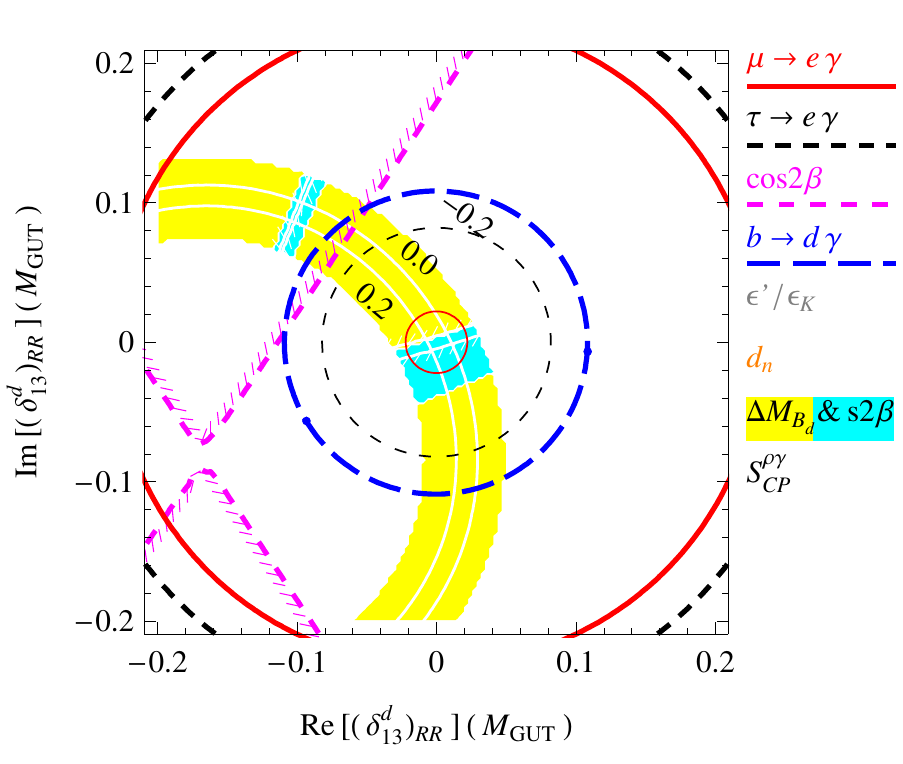}}
  \subfigure[$m_0 = 600 \GeV,\ M_{1/2} = 180 \GeV,\ \tb = 10$]{\incgr[height=63mm]{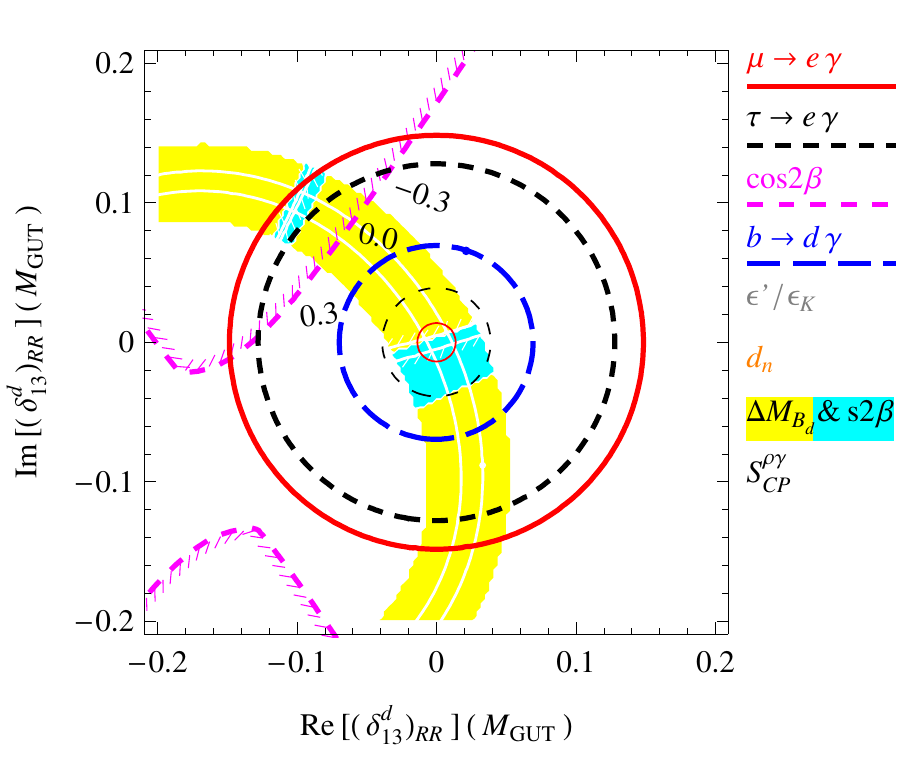}}
  \caption{Constraints on the complex plane of $\ded{13}{RR}$,
    with $\ded{ij}{LL}$ generated from
    RG running between the reduced Planck scale and the GUT scale.
    For each LFV process,
    the thick circle is the present upper bound and
    the thin circle is the prospective future bound.
    A light gray (yellow) region is allowed by $\Delta M_{B_d}$,
    given 30\% uncertainty in the $\Delta B = 2$ matrix element,
    and a gray (cyan) region is further consistent with $\sin 2 \beta$.
    The white curves running along the belt mark a possible improved
    constraint from $\Delta M_{B_d}$
    with 8\% hadronic uncertainty.
    The other white lines running across the belt
    display a measurement of $\sin 2 \beta$ at a super $B$ factory.
    Of the two sides of a $\cos 2\beta$ curve
    or a white $\sin 2 \beta$ curve,
    the excluded one is indicated by thin short lines.}
  \label{fig:13RR}
\end{figure}%
At this point we stop considering 2--3 mixing of $\fbar$,
and apply the same procedure to the 1--3 sector.
We present in Figs.~\ref{fig:13RR}, complex versions of
Figs.~\ref{fig:twoparameterscan}~(c) and (d) with the $LL$ insertions fixed at
the values listed in Table~\ref{tab:insertions}.
They exhibit constraints on $\ded{13}{RR} = \del{13}{LL}^*$.
We start over with Fig.~\ref{fig:13RR}~(a).
The light gray (yellow) belt is compatible with $\Delta M_{B_d}$,
which is further reduced by $\sin 2\beta$ into the 
gray (cyan) region.
The resulting area is completely consistent with $\bdg$.
The width of this area is comparable to the diameter of
the circle from $\teg$.
The restriction from $\meg$ is so strong that it rules out
most of the gray zone.
The $\meg$ disk in this plot is smaller than that in Fig.~\ref{fig:23RR}~(a).
The reason is that the decay amplitude is proportional to
$\del{23}{RR} \sim \lambda^2$
here, but to $\del{13}{RR} \sim \lambda^3$ there.
In a few years, improved lattice QCD should be able to
narrow the $\Delta M_{B_d}$ belt
down to the one between the two white curves,
whose width is again comparable to the diameter of the future $\teg$ disk.
If this narrowed belt is complemented by measurement of $\sin 2\beta$
at a super $B$ factory,
the combined constraint could be comparable to or stronger than
the future $\teg$ bound.
The MEG constraint is so tight that the circle
appears to be a single dot at the origin.
The radius of this circle can be looked up in Table~\ref{tab:summary}.
The dotted curves are contours of $\Srhogam$.
We present its shift that can be expected obeying other constraints
in Section~\ref{sec:otherobs}.
The rest three plots are for the cases with
(b) higher $\tb$, (c) higher $m_0$, and (d) higher $m_0$ and $\tb$,
respectively.
They can be understood in the same way as each corresponding
figure in Figs.~\ref{fig:23RR} was.
Let us stress again that with higher $m_0$,
the sensitivity of hadronic observables to the GUT scale mass insertions
is reinforced while that of LFV is weakened.
In Figs.~(c) and (d), the combination of $\Delta M_{B_d}$ and $\sin 2\beta$
essentially determines the viable areas.
This trend is expected to be maintained by a super $B$ factory.
One can notice that
$d_n$ in Figs.~(c) and (d), and $\epsilon'/\epsilon_K$ in Fig.~(d),
begin to be visible due to increased $m_0$.
These quantities are susceptible to the imaginary parts of
$\ded{13}{RR}\ded{33}{RL}\ded{31}{LL}$ and
$\ded{13}{RR}\ded{33}{RL}\ded{32}{LL}$, respectively,
although they are not playing important roles here.
Figs.~\ref{fig:13RR}~(b) and (d) show that the amplitude of
$\bdg$ is enhanced by higher $\tb$.
One can see the reason replacing $s$ by $d$ in Fig.~\ref{fig:bsg}~(b).

For the sake of completeness, we report restrictions on
the $\mathbf{10}$ sector mixings as well, which can be represented by
$\ded{ij}{LL} = \del{ij}{RR}^*$.
In Figs.~\ref{fig:23LL}, we examine the 2--3 mixing.
Comparing them with Figs.~\ref{fig:23RR},
one can notice that
\bsbsbar\ mixing is not as restrictive
on $\ded{23}{LL}$ as $\ded{23}{RR}$.
There,
the gluino loop contribution from $\ded{23}{RR}$
to the $\Delta B = 2$ transition was enhanced by
$\ded{23}{LL}$ from radiative correction.
By contrast, $\ded{23}{LL}$ here is not reinforced by $\ded{23}{RR}$
which is set to zero.
Nonetheless, $\Delta M_{B_s}$ and $\phi_{B_s}$ exclude part of
the plane in Figs.~\ref{fig:23LL}~(c) and (d) where
their sensitivities to the GUT scale squark mixing are maximized.
These constraints should be strengthened in the future.
The white curves with short thin lines attached to them
mark an improved $\phi_{B_s}$ measurement at LHCb.
They appear in all the four cases.
The other white curves, appearing in Figs.~(c) and (d), represent
the projected $\Delta M_{B_s}$ limit.
Another outstanding point is that $\bsg$ is excluding larger area of
the $\ded{23}{LL}$ plane than $\ded{23}{RR}$.
Recall that the supersymmetric diagram arising from an $LL$ mixing is added to
the SM piece since they have the same chirality structure,
while they do not interfere in the $RR$ insertion case
in Figs.~\ref{fig:23RR}.
This bound grows more stringent for higher $\tb$ as is evident
from plots in the right column.
The LFV constraints are noticeably weaker here than in Figs.~\ref{fig:23RR}.
Concerning $\tmg$, this is because
the decay is dominated by neutralino loop here,
but by chargino loop there.
The chargino loop, if present,
generically has higher effectiveness per mass insertion size,
than the neutralino loop.
One can find that $\meg$ also occurs.
It is caused by a neutralino loop graph proportional to
$\del{23}{RR} \del{31}{RR}$ with RG-induced $\del{31}{RR}$.
However, it is not strengthened by the factor $m_\tau / m_\mu$,
which accounts for the lower branching ratio than in Figs.~\ref{fig:23RR}.
Despite being moderate,
the present and future LFV bounds are still disallowing portions
of the parameter space.
The dotted contours show $\ACPbsgam$, the direct $CP$ asymmetry in
$B \rightarrow X_s \gamma$.
The numerical value of its variation is shown in Table~\ref{tab:monitored},
together with that of another related $CP$ asymmetry, $\ACPbsdgam$.

Now, we switch to the HFAG fit of the phase of \bsbsbar\ mixing.
In Figs.~\ref{fig:23RR}~(a) and (b), we cannot find a point
which falls within the 90\% CL range of $\phi_{B_s}$,
even if we allow for an $\order(1)$ squark mixing.
Favored regions appear in Figs.~(c) and (d), where
hadronic processes are enhanced.
As those regions involve a large 2--3 mixing of
left handed down-type squarks, they are likely to give a large modification
to $\bsg$, in particular for high $\tb$.
In Fig.~(d), one can notice that a substantial part of the zone of
$\ded{23}{LL}$,
needed to fit $\phi_{B_s}$, may conflict with $\SphiK$.
This conflict also grows more serious with increasing $\tb$.
In an attempt to account for the negative value of $\phi_{B_s}$
with an $LL$ mixing,
one could have a bigger hope,
given a large mixing, higher $m_0$, and low $\tb$.
Even if this scenario is realized,
$\tmg$ and $\meg$ will be hard to observe even at a super $B$ factory or MEG\@.

\begin{figure}
  \centering
  \subfigure[$m_0 = 220 \GeV,\ M_{1/2} = 180 \GeV,\ \tb = 5$]{\incgr[height=63mm]{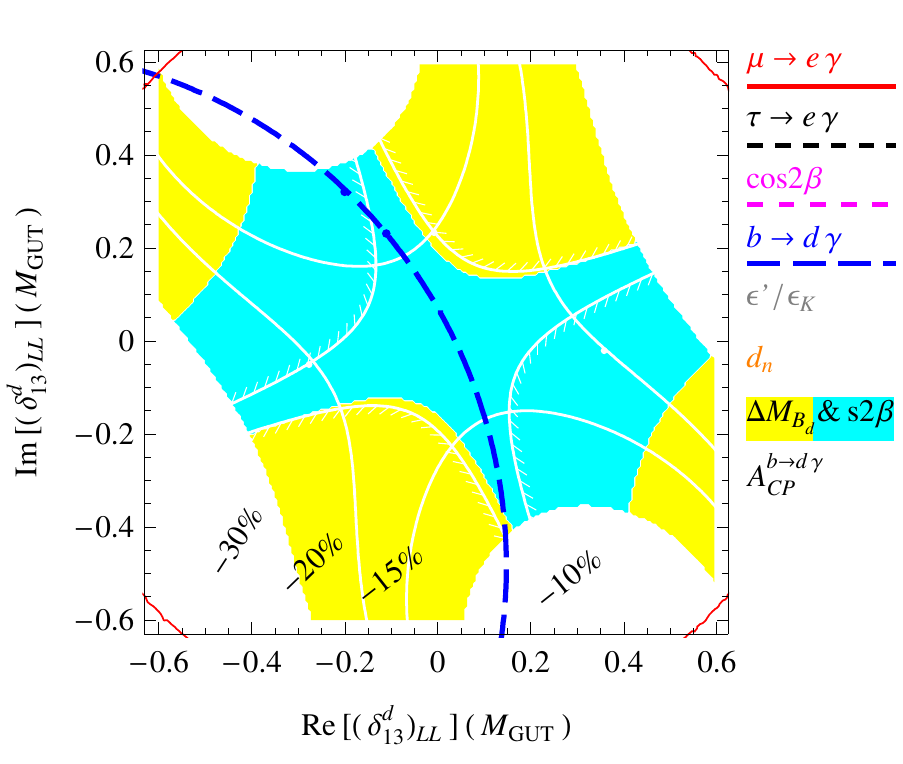}}
  \subfigure[$m_0 = 220 \GeV,\ M_{1/2} = 180 \GeV,\ \tb = 10$]{\incgr[height=63mm]{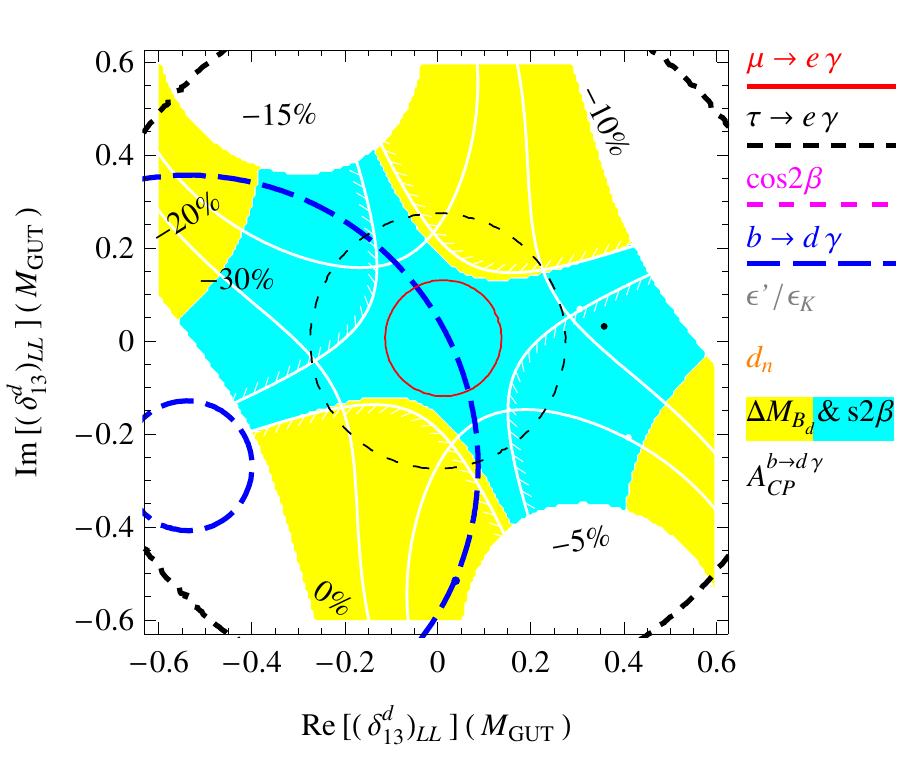}}
\\
  \subfigure[$m_0 = 600 \GeV,\ M_{1/2} = 180 \GeV,\ \tb = 5$]{\incgr[height=63mm]{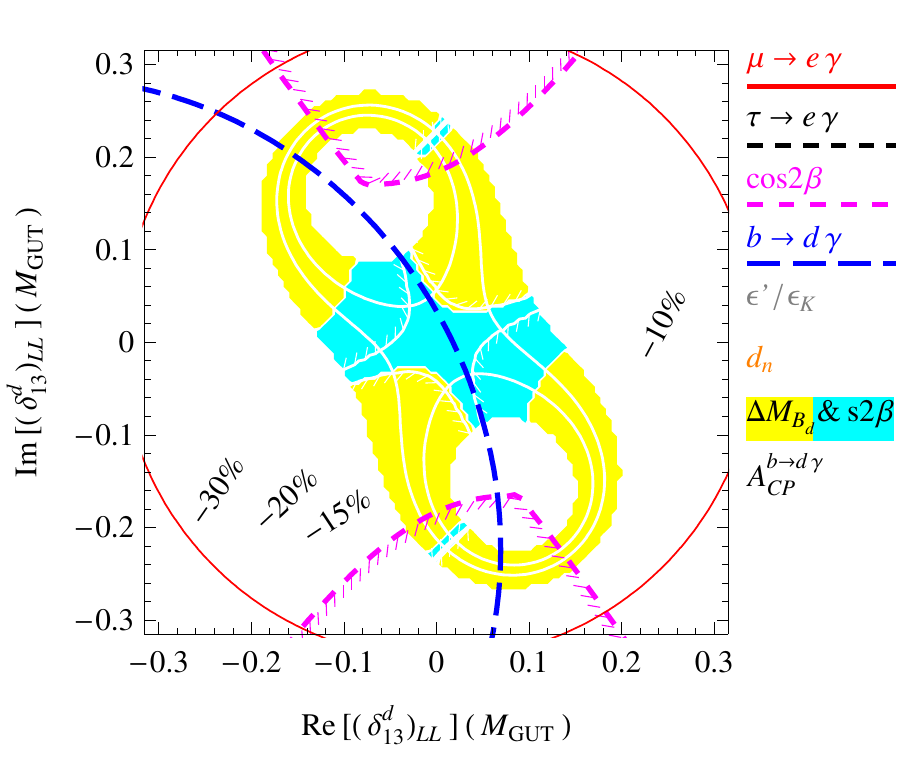}}
  \subfigure[$m_0 = 600 \GeV,\ M_{1/2} = 180 \GeV,\ \tb = 10$]{\incgr[height=63mm]{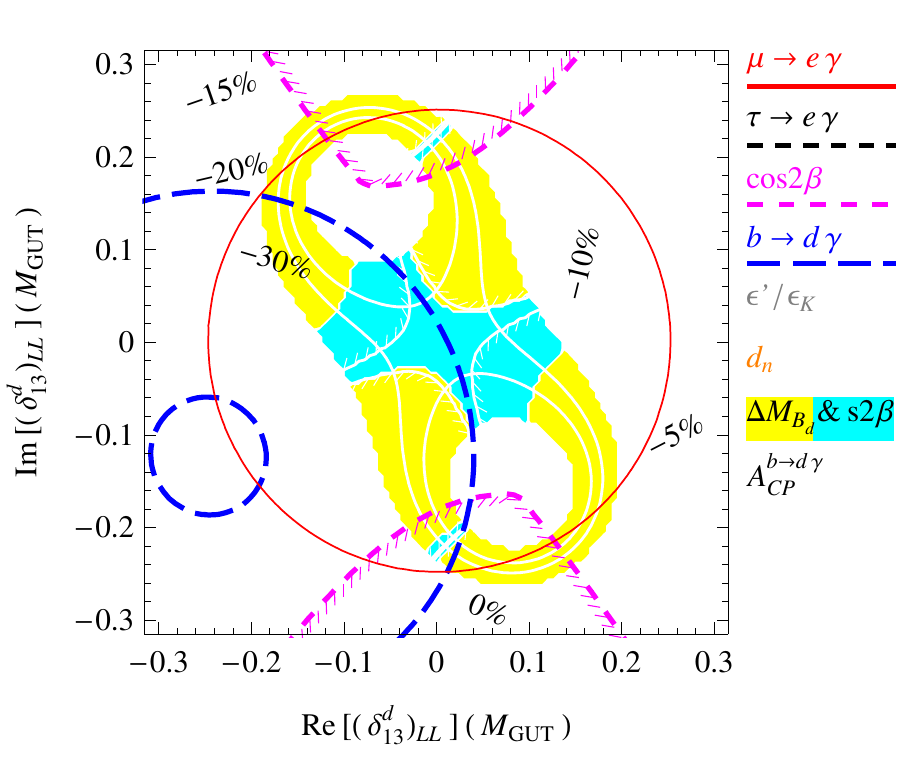}}
  \caption{Constraints on the complex plane of $\ded{13}{LL}$,
    with $\ded{12}{LL}$ and $\ded{23}{LL}$ generated from
    RG running between the reduced Planck scale and the GUT scale.
    For each LFV process,
    the thick circle is the present upper bound and
    the thin circle is the prospective future bound.
    A light gray (yellow) region is allowed by $\Delta M_{B_d}$,
    given 30\% uncertainty in the $\Delta B = 2$ matrix element,
    and a gray (cyan) region is further consistent with $\sin 2 \beta$.
    The white curves without short thin lines attached to them
    mark a possible improved
    constraint from $\Delta M_{B_d}$
    with 8\% hadronic uncertainty.
    The white curves with short thin lines attached to them
    display a measurement of $\sin 2 \beta$ at a super $B$ factory.
    Of the two sides of a $\cos 2\beta$ curve
    or a white $\sin 2 \beta$ curve,
    the excluded one is indicated by the thin short lines.}
  \label{fig:13LL}
\end{figure}

Finally, we proceed to the exclusion plots on the complex plane of
$\ded{13}{LL} = \del{13}{RR}^*$ in Figs.~\ref{fig:13LL}.
Let us compare these with those in Figs.~\ref{fig:13RR}.
A gray (cyan) region here is larger.
A significant portion of a gray zone 
is cut out by $\bdg$ \cite{Ko:2002ee}.
The LFV circles are significantly bigger.
Each of the above facts can be explained in a parallel fashion
as we did in the previous paragraph.
The dotted contours are values of $\ACPbdgam$.
Its discussion
will follow in a later part.
Note that there are cases where the future LFV
data may kill part of the area that is compatible with
$B$ physics measurements.
An example is shown in Fig.~\ref{fig:13LL}~(b)
with lower $m_0$ and higher $\tb$.
Yet, constraints mostly come from the hadronic sector.

We finish this subsection with a remark on the sizes of
the allowed regions shown in the preceding figures.
We use mass insertion parameters at the GUT scale as
the horizontal and vertical axes.
Therefore, one should be careful in comparing a plot in this paper with
one from another work, when the latter is using mass insertions
at the weak scale.
If the weak scale variable is a squark mass insertion,
one should convert our plots using~\eqref{eq:squarkMIdilution}
beforehand.

\subsection{Non-renormalizable operators and leptonic constraints}
\label{sec:nrlfv}

In the numerical analysis of the previous subsection, we have been employing
the naive relations~\eqref{eq:naivecorr}.
Now, we should discuss how the results will change
if we relax this simplification and generalize
the correlation of mass insertions to~\eqref{eq:mirelations}.
One could easily guess that the one-to-one correspondence between
a hadronic and a leptonic channel should be disturbed.
Yet, it is not completely broken as will be shown below.
Tau decay modes still limit 1--3 and 2--3 mixings of squarks
of either chirality, albeit to a reduced extent.
Similarly, $\meg$ remains a constraint on the $RR$ insertions.

Let us think about how a tau decay bound should be modified.
We first focus on $RR$ insertions, and then on $LL$.
If there is an $RR$ mixing,
a tau decay amplitude is dominated by the chargino loop
which is proportional to $\del{13}{LL}$ for $\teg$
or $\del{23}{LL}$ for $\tmg$.
Neglecting the $\order(\xi)$ term in~\eqref{eq:mirelationRR},
one has
\begin{equation}
  \label{eq:dela3LL}
  \del{a3}{LL} =
  [U_L]_{ab}\, \ded{b3}{RR}^* \, [U_L]_{33}^* +
  \order(\cos^2\!\beta\, \deltad_{RR}) , \quad
  a, b = 1, 2 ,
\end{equation}
using~\eqref{eq:1323mixing},
the smallness of mixing of the third family with the other two.
[By the same token as for (\ref{eq:dela3RR}), there can be
another term $[U_L]_{a3}\, \ded{33}{RR} \, [U_L]_{33}^*
\sim
1.5\times 10^{-5} / \cb + 0.17 \,\lambda_N^2\cb$
where we use the $(3,3)$ component of~(\ref{eq:Deltagm2F}) for its estimation.
For small neutrino Yukawa couplings,
which we assume in the numerical analysis,
this term is negligible even compared to the smallest upper bound
that can be found in Table~\ref{tab:summary}.
For large $\lambda_N$, this should be an uncertainty
in relating squark and slepton mixings,
apart from that stemming from running below $\MGUT$.]
The mixing between the first and the second families, parametrized by
$[U_L]_{ab}$, is not limited to be small.
For instance, consider the case where $\ded{23}{RR}$ is nonzero
while $\ded{13}{RR}$ is zero, as in Figs.~\ref{fig:23RR}.
Here, $\tmg$ provides a significant constraint on $\ded{23}{RR}$
if $U_L$ is a unit matrix.
Otherwise, it might happen that the association of $\del{23}{LL}$
with $\ded{23}{RR}$ is weakened by the factor $[U_L]_{22}$, or
in the worst case, is completely broken for $[U_L]_{22} = 0$.
Although $\tmg$ does not occur
in this extreme situation, $\teg$ does since $\ded{23}{RR}$ gives rise to
it through $\del{13}{LL}$ due to the approximate unitarity of $[U_L]_{ab}$.
This argument, for a general $[U_L]_{ab}$, can be summarized in the form,
\begin{equation}
  \label{eq:deltau}
  |\del{13}{LL}|^2 + |\del{23}{LL}|^2 \approx
  |{\ded{13}{RR}}|^2 + |{\ded{23}{RR}}|^2 +
  \order[\cos^2\!\beta\, (\deltad_{RR})^2] ,
\end{equation}
which determines $B(\tau \rightarrow (e + \mu)\,\gamma)$.
The mass insertions appearing above are all at the GUT scale.
Note that the current experimental bounds on $B(\tmg)$
and $B(\teg)$ differ only by a factor of 2.4.
Therefore, once one combines these two,
one can always give an upper bound on
each of $\ded{23}{RR}$ and $\ded{13}{RR}$, almost independent of $U_L$.
The error caused by non-vanishing 1--3 or 2--3 mixing in $U_L$,
is diminished below 10\% even for $\tb$ as low as 3.
If one wants to apply this conservative constraint to
the case of Figs.~\ref{fig:23RR},
the radius of each thick $\tmg$ circle should be enlarged by a factor of 1.9.
The thick $\teg$ circles in Figs.~\ref{fig:13RR}
should be expanded by a factor of 1.2.
Similarly, the future bounds can be modified:
multiply each by $\sqrt{2}$.
Even in this case, tau decays remain severe constraints on
sfermion mixings.

The same prescription can be applied to the tau decay bound on an $LL$ mixing.
Except that the amplitude is dominated by a neutralino loop,
we can repeat the above line of reasoning with $L$ and $R$ exchanged.
In this case, a possible additional term in~\eqref{eq:dela3RR}
arising from $\ded{33}{LL}$, discussed in Section~\ref{sec:running},
is negligible relative to an upper limit from $\tmg$ or $\teg$
shown in Figs.~\ref{fig:23LL} and \ref{fig:13LL}.
One can obtain a region permitted by
$\tau \rightarrow (e + \mu)\,\gamma$
in Figs.~\ref{fig:23LL}, multiplying the radius of a thick $\tmg$
circle by 1.9.
The expansion factor for Figs.~\ref{fig:13LL} is 1.2.
Again, each future bound should be multiplied by $\sqrt{2}$.

Unlike the tau decay modes, $\meg$ is more involved,
and the following method is applicable only to an $RR$ insertion.
The dominant contribution comes from the triple insertion graph
in Fig.~\ref{fig:megtriple}.
Including the diagram with opposite chirality structure,
we find that the decay rate is proportional to
\begin{equation}
  \label{eq:megpropto}
  \begin{aligned}
    d \equiv |\del{13}{RR} \del{32}{LL}|^2 + |\del{13}{LL} \del{32}{RR}|^2 .
  \end{aligned}
\end{equation}
The $\meg$ data supplies an upper limit on this quantity.
One can use~\eqref{eq:deltau} to show that
\begin{equation}
  \label{eq:dinequality}
  d \gtrsim
  \min \{ |\del{13}{RR}|^2, |\del{23}{RR}|^2 \} \cdot
  [ |{\ded{13}{RR}}|^2 + |{\ded{23}{RR}}|^2 ] ,
\end{equation}
ignoring the term suppressed by $\cos^2\!\beta$.
In contrast to $\tau \rightarrow (e + \mu)\,\gamma$,
$\meg$ depends on
the new pivotal factors, $\del{13}{RR}$ and $\del{23}{RR}$.
Ignoring the non-renormalizable operators, we had their values
equal to those of $\ded{13}{LL}^*$ and $\ded{23}{LL}^*$
in Table~\ref{tab:insertions}, respectively.
As to how $\del{13}{RR}$ and $\del{23}{RR}$ change after
the non-renormalizable operators are turned on,
there are three logical possibilities:
(a) each value remains at the same order of magnitude;
(b) either is very small and the other is not;
(c) both are vanishingly small.
In Case~(a), one can use~\eqref{eq:dinequality}
in order to translate the upper limit on $d$ to those on
$\ded{13}{RR}$ and $\ded{23}{RR}$, nearly independent of $U_L$.
We have seen that both $\del{13}{RR}$ and $\del{23}{RR}$ are
at least of the same order as $\ded{13}{LL}$ from~\eqref{eq:delRRCKM}---%
otherwise, they should belong to Case~(b) or (c).
Thus, the $U_L$-independent upper bound on each of
$\ded{13}{RR}$ and $\ded{23}{RR}$, should be
given by a $\meg$ ring in Figs.~\ref{fig:23RR}.
That is, Figs.~\ref{fig:23RR} are not modified even with this conservative
interpretation, 
while the $\meg$ circles in Figs.~\ref{fig:13RR}
should be replaced by those in Figs.~\ref{fig:23RR}.
In Case~(b), the bound inevitably depends on $U_L$.
As above, consider the scenario where $\ded{23}{RR}$ is non-vanishing
while $\ded{13}{RR}$ vanishes.
In addition, suppose that $\del{23}{RR}$, for example,
happens to be highly suppressed.
Then, \eqref{eq:dela3LL} and \eqref{eq:megpropto} lead to
\begin{equation}
  d \approx |\del{13}{RR}|^2|\ded{23}{RR}|^2|[U_L]_{22}|^2 .
\end{equation}
The branching ratio scales like $|[U_L]_{22}|^2$.
Therefore, a $\meg$ circle in Fig.~\ref{fig:23RR}
should be enlarged by the factor $1/|[U_L]_{22}|$.
However, we have learned in Section~\ref{sec:running} that
Case~(b) is not realized unless the mixing angle in $[U_R]_{ab}$ is fine-tuned.
In Case~(c),
which requires a conspiracy of
$\lambdau$, $\lambdad$, $h_1$, $h_2$, $f_1$, and $f_2$
in~\eqref{eq:wnr}, as well as the soft terms,
$\meg$ does not serve as a constraint.

In the last part of Section~\ref{sec:running}, we discussed
consequences of large neutrino Yukawa couplings
assuming $U_L$ to be an identity matrix.
We considered two cases: one where neutrino Yukawa couplings are fixed, and
the other where boundary condition at $\Mgrav$ is fixed
at a universal set of values.
Here, let us examine how those results change
if we relax the condition on $U_L$.
For the first case, we include $l_{ij}$ into~\eqref{eq:deltau} to obtain
\begin{equation}
  |{\ded{13}{RR}}|^2 + |{\ded{23}{RR}}|^2 \approx
  |\del{13}{LL} - l_{13}|^2 + |\del{23}{LL} - l_{23}|^2 ,
\end{equation}
where $\ded{ij}{RR}$ and $\del{ij}{LL}$ are at $\MGUT$ and $\MSUSY$,
respectively.
Unless $l_{ij}$ is small enough compared to the bound on
$\del{ij}{LL} = \ded{ij}{RR}^*$
presented in the previous subsection,
the limit on the left hand side is appreciably weakened.
Note that a model with $l_{ij}$ that large
is likely to be ruled out by LFV data.
The second case is more promising.
One can extend~\eqref{eq:delMGUTdelMSUSY}
in the style of~\eqref{eq:deltau},
to have
\begin{equation}
  |{\ded{13}{RR}}|^2 + |{\ded{23}{RR}}|^2 \approx
  \Bigl( \frac{\alpha}{1 + \alpha} \Bigr)^2 \times
  [ |\del{13}{LL}|^2 + |\del{23}{LL}|^2 ] .
\end{equation}
Therefore, the upper bounds on $\ded{23}{RR}$ and $\ded{13}{RR}$
attained from~\eqref{eq:deltau},
are further scaled down by $\alpha/(1 + \alpha)$.

Recently, an alternative approach to settling down the uncertainties
posed by the non-renormalizable operators has been reported
\cite{Borzumati:2007bn}.
Their work in progress makes use of dependence of the proton lifetime
on the coefficients of the operators \cite{Emmanuel-Costa:2003pu}
in order to find a pattern among them.
We would say that
our strategy is more generic in the sense that
it relies only on the condition that the non-renormalizable operators are
Planck-suppressed, although it may not be as predictive as their
anticipated outcome.

\subsection{Summary of bounds}
\label{sec:bounds}

\TABLE{
  \renewcommand{\arraystretch}{1.1}
  \begin{tabular}{ccccc}
    \hline
    Mixing & \multicolumn{2}{c}{Fig.} & Present & Future
    \\
    \hline
    \multirow{4}*{$|\ded{23}{RR} (\MGUT)|$} &
    \multirow{4}*{\ref{fig:23RR}} & (a)
    & $7.6\times 10^{-2}$--$1.4\times 10^{-1}$
    & $1.3\times 10^{-2}$
    \\ & & (b)
    & $3.8\times 10^{-2}$--$7.1\times 10^{-2}$
    & $8.1\times 10^{-3}$
    \\ & & (c)
    & $1.7\times 10^{-1}$
    & $4.0\times 10^{-2}$
    \\ & & (d)
    & $8.3\times 10^{-2}$--$1.5\times 10^{-1}$
    & $3.9\times 10^{-2}$--$4.3\times 10^{-2}$
    \\
    \hline
    \multirow{4}*{$|\ded{13}{RR} (\MGUT)|$} &
    \multirow{4}*{\ref{fig:13RR}} & (a)
    & $2.7\times 10^{-2}$--$1.4\times 10^{-1}$
    & $2.5\times 10^{-3}$--$1.2\times 10^{-2}$
    \\ & & (b)
    & $1.6\times 10^{-2}$--$7.0\times 10^{-2}$
    & $1.5\times 10^{-3}$--$7.3\times 10^{-3}$
    \\ & & (c)
    & $4.7\times 10^{-2}$
    & $1.1\times 10^{-2}$
    \\ & & (d)
    & $5.0\times 10^{-2}$
    & $1.1\times 10^{-2}$
    \\
    \hline
    \multirow{4}*{$|\ded{23}{LL} (\MGUT)|$} &
    \multirow{4}*{\ref{fig:23LL}} & (a)
    & $\order(1)$
    & $\order(1)$
    \\ & & (b)
    & $0.6$--$\order(1)$
    & $0.3$--$0.4$
    \\ & & (c)
    & $0.7$
    & $0.3$
    \\ & & (d)
    & $0.5$
    & $0.3$
    \\
    \hline
    \multirow{4}*{$|\ded{13}{LL} (\MGUT)|$} &
    \multirow{4}*{\ref{fig:13LL}} & (a)
    & $0.6$
    & $0.3$
    \\ & & (b)
    & $0.6$
    & $0.1$--$0.3$
    \\ & & (c)
    & $0.1$
    & $0.06$
    \\ & & (d)
    & $0.1$
    & $0.06$
    \\
    \hline
  \end{tabular}
  \caption{Upper limit on the size of each mass insertion
    of down-type squarks at the GUT scale.
  The second and third columns indicate
  the values of $m_0$, $M_{1/2}$, and $\tb$,
  used in Figs.~\ref{fig:23RR}--\ref{fig:13LL}.
  Regarding an $RR$ mixing,
  if there are two numbers separated by a dash,
  the left one is for $U_L = \mathbf{1}$ and
  the right one is for $U_L \neq \mathbf{1}$ obeying~\eqref{eq:1323mixing}.
  If the two numbers are the same, it is written only once.
  We do the same for an $LL$ mixing on which
  the alignment condition is given through $U_R$ instead of $U_L$.
  For a general $U_R$, we drop the $\meg$ constraint
  as we do not have a systematic way to impose it.}
  \label{tab:summary}
}
The restrictions on down-type squark mixings at the GUT scale,
graphically shown in Section~\ref{sec:regions},
are condensed in a numerical form in
Table~\ref{tab:summary}.
Each number is the maximum distance of a point from the origin
on the corresponding figure that satisfies all the constraints considered
in the present work.
As for $\phi_{B_s}$,
we use the D\O\ result,
which is marked in gray (cyan) in Figs.~\ref{fig:23RR} and \ref{fig:23LL}.
We would be left with no solution in many cases if we used the HFAG fit
(which would be a very interesting outcome on its own
\cite{Dutta:2008xg,Hisano:2008df,Parry:2007fe,lfvutfit}).
In order to estimate the power of $\phi_{B_s}$ measurement at LHCb,
we suppose that its future central value will coincide with the SM prediction.
We make the same supposition about $\sin 2\beta$.

Those upper bounds are subject to change
of parameters or scheme of uncertainty treatment.
A variation may also be caused by
choosing $2\times 10^{-9}$ instead of $10^{-8}$
as the reach of $\teg$ and $\tmg$ searches at a super $B$ factory.
In particular,
the strength of a LFV constraint depends on $U_L$ and $U_R$.
We take into account the $U_L$ dependence of a maximal $RR$ insertion
using the method described in the previous subsection.
As for $\meg$, we suppose Case~(a) therein, i.e.\
we do not envisage a fine tuning among contributions to
$\del{13}{RR}$ or $\del{23}{RR}$.
If a LFV restriction is important,
relaxing the assumption of $U_L = \mathbf{1}$ increases
the upper limit of the given insertion.
Concerning the limit on an $LL$ insertion, 
we follow the same procedure to evaluate the dependence of a tau channel
on $U_R$, while we keep $\meg$ only for $U_R = \mathbf{1}$.
Even if $U_R$ is unity, however, it turns out that
the leptonic data does not cause a big additional reduction
in the bounds set by the hadronic inputs,
under the conditions considered in this work.
A lepton sector constraint should be looser if we allow for a different $U_R$.
Therefore, the quoted numbers are not greatly influenced by
a change of $U_R$.

\subsection{Possible alterations in observables}
\label{sec:otherobs}

With the region of each mass insertion
obtained in Section~\ref{sec:regions},
we estimate a possible difference of an affected observable
from its SM value.
The result is summarized in Table~\ref{tab:monitorresult}.
\TABLE{
  \renewcommand{\arraystretch}{1.1}
  \begin{tabular}{cccccc}
    \hline
    Deviation & \multicolumn{2}{c}{\multirow{2}*{Fig.}} &
    \multicolumn{1}{c}{\multirow{2}*{Present}} &
    \multicolumn{1}{c}{\multirow{2}*{Future}} & Future 
    \\
    Mixing & & & & & precision
    \\
    \hline
    & \multirow{4}*{\ref{fig:23RR}} & (a)
    & $0.04$--$0.07$
    & $0.007$
    & \multirow{4}*{0.02}
    \\ $\bigl|\Delta\SKstargam\bigr|$ & & (b)
    & $0.04$--$0.07$
    & $0.007$
    \\ $\ded{23}{RR}$ & & (c)
    & $0.18$
    & $0.04$
    \\ & & (d)
    & $0.16$--$0.26$
    & $0.07$--$0.08$
    \\
    \hline
    & \multirow{4}*{\ref{fig:13RR}} & (a)
    & $0.06$--$0.30$
    & $0.006$--$0.03$
    & \multirow{4}*{0.10}
    \\ $\bigl|\Delta\Srhogam\bigr|$ & & (b)
    & $0.06$--$0.28$
    & $0.006$--$0.03$
    \\ $\ded{13}{RR}$ & & (c)
    & $0.21$
    & $0.05$
    \\ & & (d)
    & $0.39$
    & $0.09$
    \\
    \hline
    & \multirow{4}*{\ref{fig:13RR}} & (a)
    & $0.06$--$0.28$
    & $0.006$--$0.03$
    \\ $\bigl|\Delta\SBsKsgam\bigr|$ & & (b)
    & $0.06$--$0.28$
    & $0.006$--$0.03$
    \\ $\ded{13}{RR}$ & & (c)
    & $0.17$
    & $0.03$
    \\ & & (d)
    & $0.32$
    & $0.05$
    \\
    \hline
    & \multirow{4}*{\ref{fig:23LL}} & (a)
    & $1.3$
    & $1.3$
    & \multirow{4}*{0.4}
    \\ $\bigl|\Delta\ACPbsgam\bigr|$\,(\%) & & (b)
    & $1.9$--$2.3$
    & $1.0$--$1.4$
    \\ $\ded{23}{LL}$ & & (c)
    & $3.3$
    & $1.7$
    \\ & & (d)
    & $5.2$
    & $2.8$
    \\
    \hline
    & \multirow{4}*{\ref{fig:23LL}} & (a)
    & $1.3$
    & $1.3$
    & \multirow{4}*{0.6}
    \\ $\bigl|\Delta\ACPbsdgam\bigr|$\,(\%) & & (b)
    & $1.8$--$2.2$
    & $0.9$--$1.3$
    \\ $\ded{23}{LL}$ & & (c)
    & $3.2$
    & $1.6$
    \\ & & (d)
    & $5.1$
    & $2.7$
    \\
    \hline
    & \multirow{4}*{\ref{fig:13LL}} & (a)
    & $16$
    & $7$
    \\ $\bigl|\Delta\ACPbdgam\bigr|$\,(\%) & & (b)
    & $57$
    & $5$--$15$
    \\ $\ded{13}{LL}$ & & (c)
    & $7$
    & $3$
    \\ & & (d)
    & $15$
    & $6$
    \\
    \hline
  \end{tabular}
  \caption{Maximal departure of each observable from its
    SM value given the present and the future constraints.
  The second and third columns indicate
  the plot on which we calculate the observable.
  Of the two deviations separated by a dash in a cell,
  the left one is for $U_L = \mathbf{1}$ and
  the right one is for $U_L \neq \mathbf{1}$
  obeying~\eqref{eq:1323mixing},
  for the first three $CP$ asymmetries.
  Those two types of deviations should be regarded as the same
  if only one is written.
  For the rest, the alignment condition is
  given through $U_R$ instead of $U_L$.
  For a general $U_R$, we drop the $\meg$ constraint
  as we do not have a systematic way to impose it.}
  \label{tab:monitorresult}
}%
Four of them have been already
displayed as contours on each figure indicated in the table.
Note that what has been shown as contours is the value of the observable,
not the deviation from the SM prediction.
We use the same set of constraints as in Section~\ref{sec:bounds}.

Under the present conditions,
there are still $CP$ asymmetries that might potentially have a discrepancy
bigger than the precision attainable at a super $B$ factory.
They are $\SKstargam$, $\Srhogam$, $\ACPbsgam$, and $\ACPbsdgam$.
They show larger possible alterations for higher $m_0$, while
$\ACPbdgam$ doesn't follow this tendency.
Being hadronic observables, their sensitivity to the GUT scale
flavor violation is amplified for higher $m_0$,
as was explained in Section~\ref{sec:regions},
although they are more severely restricted by
other quark sector processes for the same reason.
As we did for Table~\ref{tab:summary},
we take account of uncertainties due to a misalignment between
quarks and leptons of the lighter two families.
In this case, we obtain the values after the dash signs, which
can be larger than the estimates for perfect alignment.

We repeat the same task with the prospective future inputs.
One may expect the presented deviations,
provided that no constraint is seriously violated in a future experiment.
With lower $m_0$, $\SKstargam$ and $\Srhogam$ will not
show a signature detectable at a super $B$ factory,
even if quark--lepton misalignment is allowed,
while $\ACPbsgam$ and $\ACPbsdgam$ might reveal a hint.
With higher $m_0$, search for a supersymmetric effect in $\SKstargam$
becomes feasible as well.

In the case with the $\ded{13}{LL}$ mixing,
its effect on $\ACPbsdgam$ is negligible
so that the variation is at most about $0.5\%$,
because the channel $B \rightarrow X_{s+d} \gamma$ is dominated by
$B \rightarrow X_s \gamma$.
We include $\SBsKsgam$ in the table as well for reference.

Among the $CP$ asymmetries mentioned above,
$\SKstargam$ and $\Srhogam$ are sensitive to $RR$ mixings of squarks,
and thus are closely related to LFV\@.
Recall that $RR$ mixings give rise to much higher LFV rates than $LL$,
as we have seen in Section~\ref{sec:regions}.
This motives us to
look into allowed ranges of those two $CP$ asymmetries
as functions of LFV branching ratios.

\begin{figure}
  \centering
\subfigure[$m_0 = 220 \GeV,\ M_{1/2} = 180 \GeV,\ \tb = 5$]{\incgr[width=72mm]{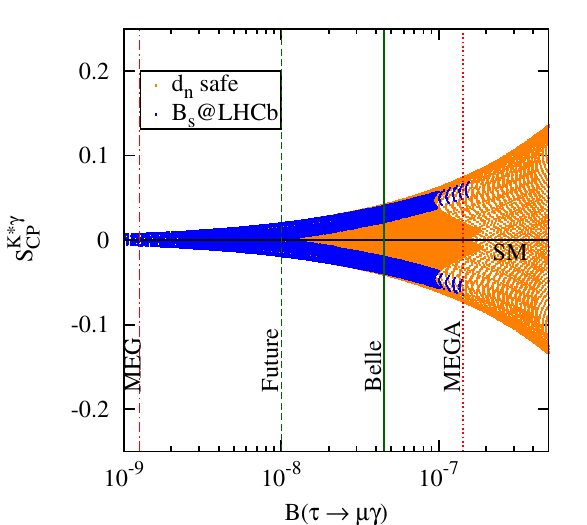}}
\subfigure[$m_0 = 220 \GeV,\ M_{1/2} = 180 \GeV,\ \tb = 10$]{\incgr[width=72mm]{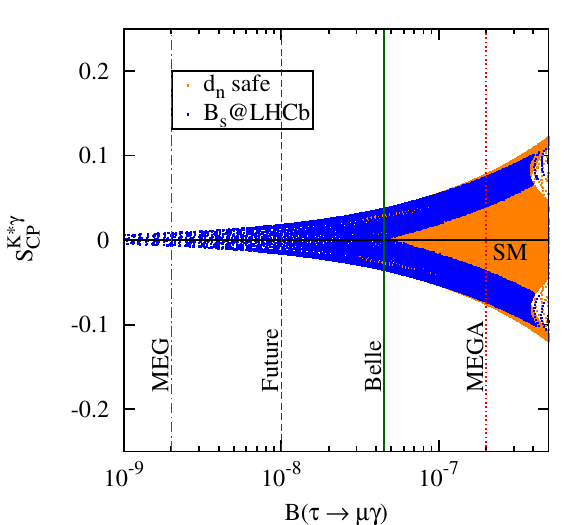}}
\\
\subfigure[$m_0 = 600 \GeV,\ M_{1/2} = 180 \GeV,\ \tb = 5$]{\incgr[width=72mm]{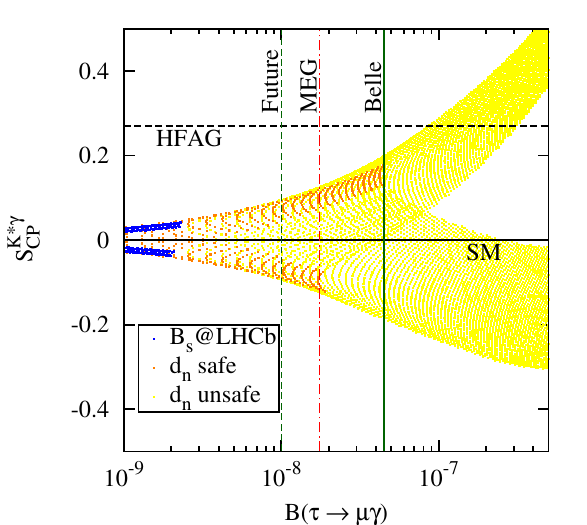}}
\subfigure[$m_0 = 600 \GeV,\ M_{1/2} = 180 \GeV,\ \tb = 10$]{\incgr[width=72mm]{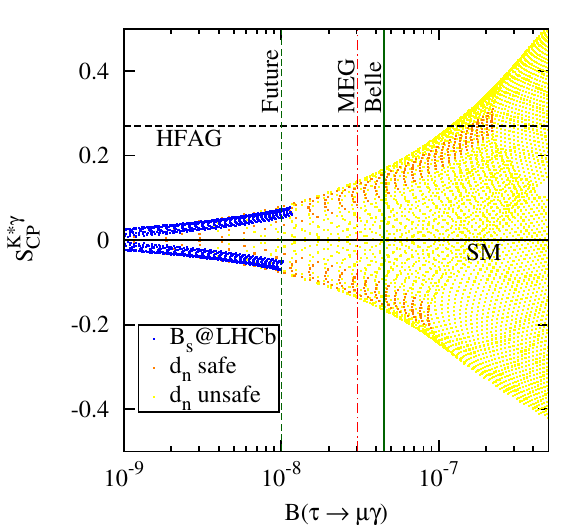}}
  \caption{Correlation between $\SKstargam$ and $B(\tmg)$
    obtained by varying $\ded{23}{RR}$,
    with $\ded{ij}{LL}$ generated from
    RG running between the reduced Planck scale and the GUT scale.
    A light gray (yellow) point is disfavored by neutron EDM,
    while a gray (orange) point is not, and
    a black (blue) point satisfies the future $\Delta M_{B_s}$ and
    $\phi_{B_s}$ constraints.
    The dashed horizontal line marks the 2 $\sigma$ range of $\SKstargam$,
    and its SM value is the solid horizontal line.
    The present and the future limits on $\tmg$ and $\meg$ are indicated by
    the vertical lines.}
  \label{fig:SKstargam}
\end{figure}

First, we show the correlation between $\SKstargam$ and
$B(\tmg)$ in Figs.~\ref{fig:SKstargam},
each of which results from the same set of mass insertions as
the corresponding plot in Figs.~\ref{fig:23RR}.
Every point on the figures satisfies the current $\Delta M_{B_s}$ and
$B(\bsg)$ constraints.
The upper limit on $\tmg$ from $\meg$ has been deduced from
the contours in Figs.~\ref{fig:23RR}.
In Figs.~\ref{fig:SKstargam}~(a) and (b), what restricts $\SKstargam$
at present is $\tmg$, and in the future $\meg$ at MEG should take over.
In Figs.~(c) and (d), $d_n$, in addition to $\tmg$, is playing
an important role, and the future expectation of $\SKstargam$
is determined by $\Delta M_{B_s}$ and $\phi_{B_s}$.
One can find the numerical range of $\SKstargam$
allowed in each of the four figures in Table~\ref{tab:monitorresult}.
Note that if two numbers are separated by a dash in the table,
one should take the left hand side since
the plots are for $U_L = \mathbf{1}$.
One could translate these plots to a case where $U_L$ is not fixed at unity,
following the prescription presented in Section~\ref{sec:nrlfv}:
regard the horizontal axis as $B(\tau \rightarrow (e + \mu)\,\gamma))$
instead of $B(\tmg)$, and shift the upper bounds on $\tmg$
rightward in accordance to this change, while
keeping the positions of the vertical lines for $\meg$.

\begin{figure}
  \centering
\subfigure[$m_0 = 220 \GeV,\ M_{1/2} = 180 \GeV,\ \tb = 5$]{\incgr[width=72mm]{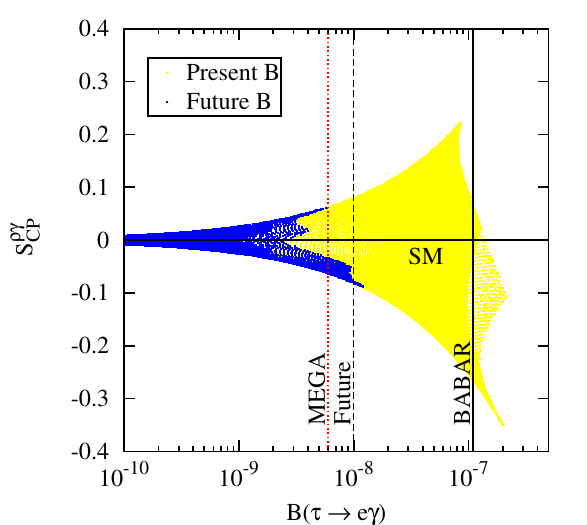}}
\subfigure[$m_0 = 220 \GeV,\ M_{1/2} = 180 \GeV,\ \tb = 10$]{\incgr[width=72mm]{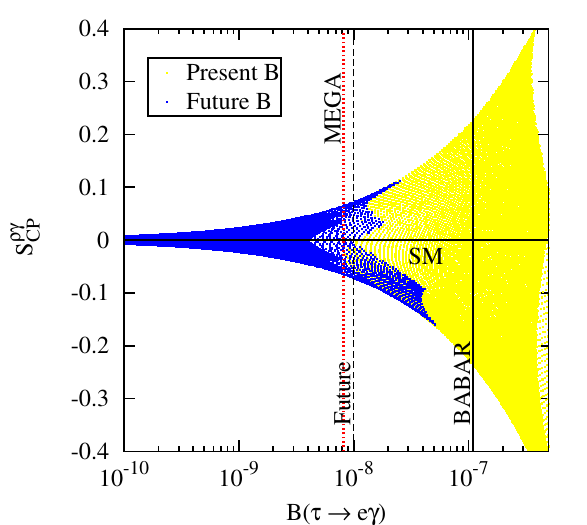}}
\\
\subfigure[$m_0 = 600 \GeV,\ M_{1/2} = 180 \GeV,\ \tb = 5$]{\incgr[width=72mm]{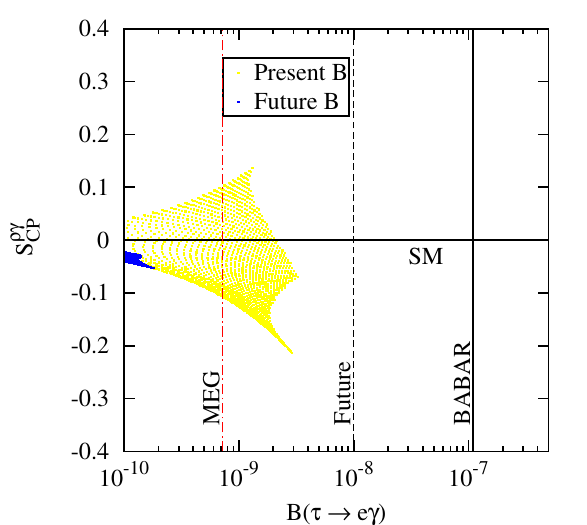}}
\subfigure[$m_0 = 600 \GeV,\ M_{1/2} = 180 \GeV,\ \tb = 10$]{\incgr[width=72mm]{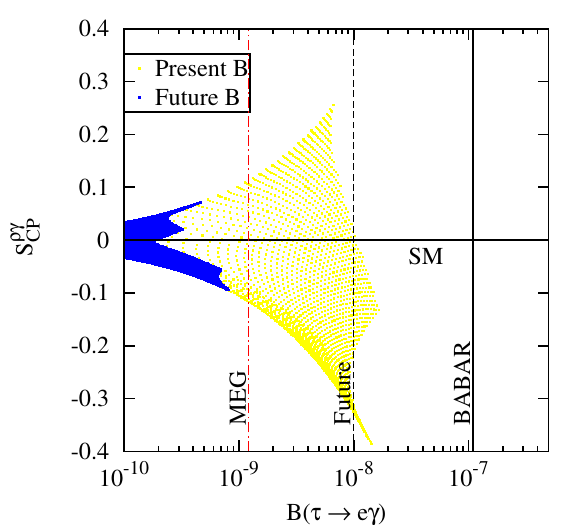}}
  \caption{Correlation between $\Srhogam$ and $B(\teg)$
    obtained by varying $\ded{13}{RR}$,
    with $\ded{ij}{LL}$ generated from
    RG running between the reduced Planck scale and the GUT scale.
    A light gray (yellow) point is consistent with all the current constraints,
    and a black (blue) point satisfies the future $\Delta M_{B_d}$ and
    $\sin 2 \beta$ constraints.
    The solid horizontal line marks the SM value of $\Srhogam$.
    The present and the future limits on $\tmg$ and $\meg$ are indicated by
    the vertical lines.
    In Figs.~(a) and (b), the MEG line is
    outside the left border of each plot.}
  \label{fig:Srhogam}
\end{figure}

Second, let us move to the correlation between $\Srhogam$ and $B(\teg)$,
displayed in Figs.~\ref{fig:Srhogam},
which correspond to the parameter space considered in Figs.~\ref{fig:13RR}.
We discard any point that is incompatible with the present data of
$\Delta M_{B_d}$, $\sin 2 \beta$, or $\cos 2 \beta$.
The upper limit on $\teg$ from $\meg$ has been inferred
as we did in the preceding paragraph.
For lower $m_0$ shown in Figs.~\ref{fig:Srhogam}~(a) and (b),
$\meg$ provides the limits on $\Srhogam$ both currently and in the future.
The MEG bound is not visible on the plane since it restricts
$B(\teg) \lesssim 7\times 10^{-11}$.
For higher $m_0$ in Figs.~(c) and (d),
possible range of $\Srhogam$ is determined by the other hadronic observables,
with little help from the lepton sector.
The way to convert these plots to those for $U_L \neq \mathbf{1}$ is
almost the same as above:
relabel the horizontal axis as $B(\tau \rightarrow (e + \mu)\,\gamma))$
instead of $B(\teg)$, and change the upper bounds on $\teg$
to those on $\tau \rightarrow (e + \mu)\,\gamma$.
A difference from the above case
is that one should also multiply the $\meg$ limit on $B(\teg)$
by $24 \sim \lambda^{-2}$.

The latest interest in the phase of \bsbsbar\ mixing leads us to
examine its modification that can be caused by new physics.
We lift the constraint on $\phi_{B_s}$ while
keeping the others used in Section~\ref{sec:bounds},
and record its variation allowed by the other bounds
in Table~\ref{tab:phis}.
\begin{table}
  \centering
  \renewcommand{\arraystretch}{1.1}
  \begin{tabular}{ccccc}
    \hline
    Deviation &
    \multicolumn{2}{c}{\multirow{2}*{Fig.}} &
    \multirow{2}*{Present} & \multirow{2}*{Future}
    \\
    Mixing & & & &
    \\
    \hline
    & \multirow{4}*{\ref{fig:23RR}} & (a)
    & $0.05$--$0.08$
    & $0.01$
    \\ $\bigl|\Delta\phi_{B_s}|$ & & (b)
    & $0.02$--$0.04$
    & $0.004$
    \\ $\ded{23}{RR}$ & & (c)
    & $0.08$
    & $0.08$
    \\ & & (d)
    & $0.05$
    & $0.05$
    \\
    \hline
    & \multirow{4}*{\ref{fig:23LL}} & (a)
    & $0.05$
    & $0.05$
    \\ $\bigl|\Delta\phi_{B_s}|$ & & (b)
    & $0.02$--$0.03$
    & $0.004$--$0.008$
    \\ $\ded{23}{LL}$ & & (c)
    & $0.57$
    & $0.33$
    \\ & & (d)
    & $0.32$
    & $0.12$
    \\
    \hline
  \end{tabular}
  \caption{Maximal departure of $\phi_{B_s}$ from its
    SM value under the present and the future constraints
    except for those on itself.
  The second and third columns indicate
  the relevant plot.
  Of the two deviations separated by a dash in a cell,
  the left one is for $U_L = \mathbf{1}$ and
  the right one is for $U_L \neq \mathbf{1}$
  obeying~\protect\eqref{eq:1323mixing},
  for the $RR$ mixing.
  Those two types of deviations should be regarded as the same
  if only one is written.
  In the case with $\ded{23}{LL}$,
  the alignment condition is given through $U_R$ instead of $U_L$.
  For a general $U_R$, we drop the $\meg$ constraint
  as we do not have a systematic way to impose it.}
  \label{tab:phis}
\end{table}
The difference between the announced central value
and the SM prediction is about $0.7$.
 From the table it appears that
cases with lower $m_0$ and/or
large $RR$ mixing (but small $LL$ mixing)
are disfavored by $\phi_{B_s}$.
In the case of $RR$ insertion with higher $m_0$,
the primary barrier is the neutron EDM
as is evident from Figs.~\ref{fig:23RR}~(c) and (d).
Let us remind the reader that this situation can be
ameliorated by multiplying $\ded{23}{LL}$ by an $\order(1)$ complex factor
at $\MGUT$.
With the $LL$ insertion and higher $m_0$,
on the other hand, Figs.~\ref{fig:23LL}~(c) and (d)
show that $\bsg$ and $\SphiK$ exclude
a major part of the region preferred by the HFAG fit, although
there are still remaining parts that are responsible for
the large difference in $\phi_{B_s}$ recorded in the table.

\begin{figure}
  \centering
\subfigure[$m_0 = 220 \GeV,\ M_{1/2} = 180 \GeV,\ \tb = 5$]{\incgr[width=72mm]{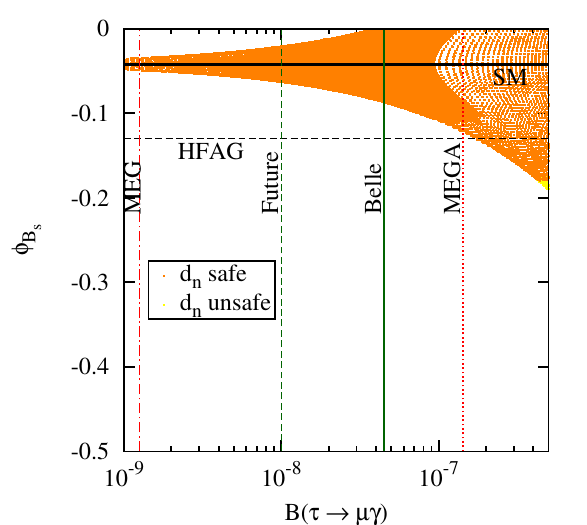}}
\subfigure[$m_0 = 220 \GeV,\ M_{1/2} = 180 \GeV,\ \tb = 10$]{\incgr[width=72mm]{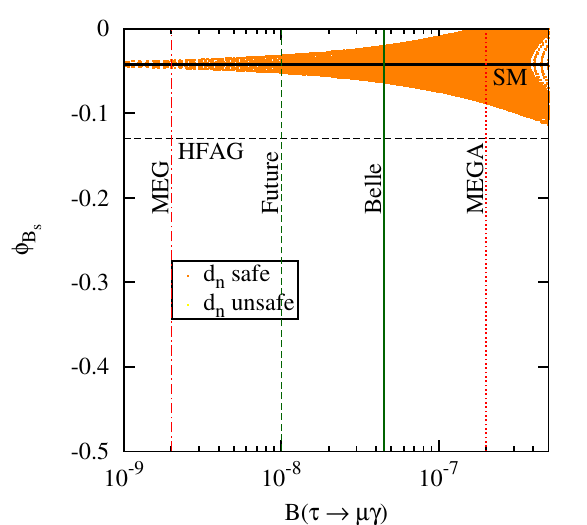}}
\\
\subfigure[$m_0 = 600 \GeV,\ M_{1/2} = 180 \GeV,\ \tb = 5$]{\incgr[width=72mm]{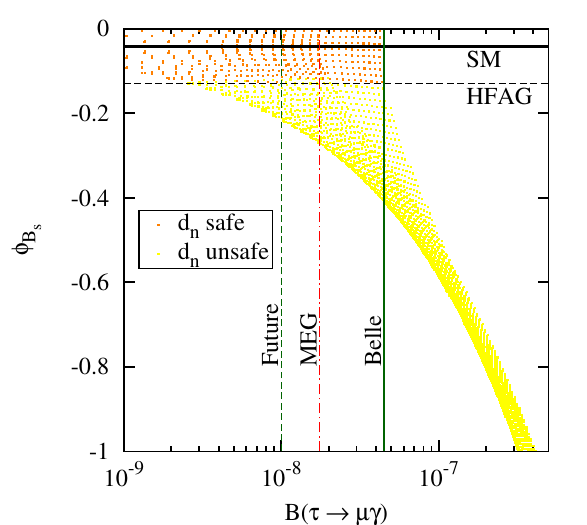}}
\subfigure[$m_0 = 600 \GeV,\ M_{1/2} = 180 \GeV,\ \tb = 10$]{\incgr[width=72mm]{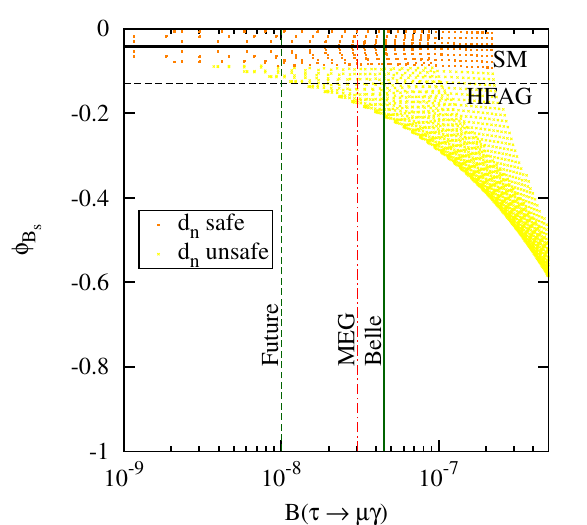}}
  \caption{Correlation between $\phi_{B_s}$ and $B(\tmg)$
    obtained by varying $\ded{23}{RR}$,
    with $\ded{ij}{LL}$ generated from
    RG running between the reduced Planck scale and the GUT scale.
    A light gray (yellow) point is disfavored by neutron EDM,
    while a gray (orange) point is not.
    The dashed horizontal line marks the 90\% CL range of $\phi_{B_s}$,
    and its SM value is the solid horizontal line.
    The present and the future limits on $\tmg$ and $\meg$ are indicated by
    the vertical lines.}
  \label{fig:phis}
\end{figure}

In Figs.~\ref{fig:phis},
we investigate how LFV constrains $\phi_{B_s}$
by means of correlation plots, focusing on
the $RR$ insertion case considered in Figs.~\ref{fig:23RR}.
At first, let us consider only the leptonic constraints.
In this case, LFV and the latest $\phi_{B_s}$ fit
are better reconciled for higher $m_0$ depicted in Figs.~(c) and (d).
Obviously, lower $\tb$ is preferable since a LFV limit gets tighter
for higher $\tb$.
However,
if one takes the neutron EDM bound seriously,
the light gray (yellow) points are discarded
while the gray (orange) points remain, and therefore
it becomes harder to account for $\phi_{B_s}$ with
an $RR$ mixing.
Remember that one can apply this result to a popular benchmark scenario
in which the soft terms at $\Mgrav$ are flavor-blind and
all the right-handed squark mixings are supposed to
originate from large neutrino Yukawa couplings,
as we discussed in the last part of Section~\ref{sec:running}.
The recipe is to multiply each LFV branching fraction by
$(1+\alpha)^2/\alpha^2$ with $\alpha$ in~\eqref{eq:alpha}.
This factor arises from the additional running of slepton masses
from $\MGUT$ down to $\MR$,
and strengthens LFV as the result.

One can be more optimistic in viewing the same correlation plots.
For example, the neutron EDM constraint may be weakened
if there is also a non-vanishing complex $LL$ mass insertion at $\Mgrav$,
or one might simply choose to ignore the constraint
due to its hadronic uncertainties.
Then, it might be that the present status of $\phi_{B_s}$ is hinting at
a LFV process occurring at a rate that can be explored in the near future.
Notice that this scenario works best when
the value of $x$ defined in~\eqref{eq:x}, is around $1/12$,
as we discussed in Section~\ref{sec:regions}.

\section{Conclusions}
\label{sec:conclusion}

We imposed hadronic and leptonic constraints
on sfermion mixing in a class of supersymmetric models
with SU(5) grand unification.
We did not particularly assume that
the sfermion mass matrices have a universal form at any scale,
but rather that any off-diagonal entry may be nonzero,
which is generically the case in gravity mediated supersymmetry breaking.
Those off-diagonal elements are encoded in the
dimensionless mass insertion parameters
in terms of which we express
experimental bounds on flavor non-universality at the GUT scale.
While fixing the gluino mass to $500\GeV$ at the weak scale,
we tried two different boundary conditions on the diagonal components
of the soft scalar mass matrix at $\MGUT$: lower $m_0 = 220\GeV$
and higher $m_0 = 600\GeV$.
We varied $\tb$ from 5 to 10 as well.
For lower $m_0$, we have found that the upper limit on
an $RR$ mixing is essentially determined by a LFV decay mode
both at present and in the near future.
This is true even when one introduces non-renormalizable terms
to accommodate the lighter down-type quark and charged lepton masses.
In particular, the apparently unrelated mode $\meg$
turns out to be remarkably sensitive to a mixing involving the third family.
This sensitivity will be much higher
with the progress of the MEG experiment.
For higher $m_0$, the situation turns the other way around
so that the hadronic constraints, such as
$B$-meson mixing and neutron EDM, dominate.
Also in the near future,
measurements at the LHCb and a super $B$ factory, with the aid of
improved lattice QCD, should be able to probe an $RR$ mixing,
with a sensitivity higher than that of a LFV experiment.
Concerning the $LL$ mixings, they are mostly restricted by
hadronic data from $B$ physics,
although LFV supplies additional information
if $m_0$ is low and $\tb$ is high.
 These findings unveil
 a nice complementarity of the quark and the lepton sector
processes showing their strengths and weaknesses,
depending on the gaugino to scalar mass ratio.
We included discussions on the consequences of the discrepancy
recently observed in the $B_s$-meson mixing phase.

\acknowledgments

We thank
Sung-Gi Kim,
Paride Paradisi,
Amarjit Soni, and
Diego Tonelli
for useful discussions and comments.
JhP acknowledges Research Grants funded jointly by the Italian
Ministero dell'Istruzione, dell'Universit\`{a} e della Ricerca (MIUR),
by the University of Padova, and
by the Istituto Nazionale di Fisica Nucleare (INFN) within the
\textit{Astroparticle Physics Project}, and the FA51 INFN Research Project,
as well as the JSPS postdoctoral fellowship program
for foreign researchers and the accompanying grant-in-aid no.\ 17.05302.
This research was supported in part by the European Community Research
Training Network UniverseNet under contract MRTN-CT-2006-035863.
The work of MY was partially supported by the grants-in-aid from the
Ministry of Education,
Science, Sports and Culture in Japan, No.\ 16081202 and No.\ 17340062.

\appendix
\section{Notations}
\label{sec:appendix}

The scalar mass terms in the soft supersymmetry breaking sector of
the minimal supersymmetric standard model are given by
\begin{equation}
  - \mathcal{L}_\mathrm{soft} \supset
  Q^\dagger m^2_Q \,Q + 
  \overline{U}^T\! m^2_U \,\overline{U}^* +
  \overline{E}^T\! m^2_E \,\overline{E}^* +
  \overline{D}^T\! m^2_D \,\overline{D}^* +
  L^\dagger m^2_L \,L ,
\end{equation}
where the uppercase letters denote the scalar components of
the SM superfields embedded in
$T$ and $\overline{F}$ as in~\eqref{eq:tenfbar}.
Consider a basis
where the down-type quark and the charged lepton Yukawa matrices
are diagonalized by superfield rotations.
The scalars in this basis, denoted by lowercase letters,
are related the above fields by~\eqref{eq:embed}.
Therefore, their mass matrices are connected to those above
by the basis change,
\begin{equation}
  m^2_q = m^2_Q, \quad
  m^2_u = U_Q \,m^2_U \,U_Q^\dagger , \quad
  m^2_e = U_R \,m^2_E \,U_R^\dagger , \quad
  m^2_d = m^2_D, \quad
  m^2_l = U_L \,m^2_L \,U_L^\dagger .
\end{equation}

Suppose that the squark and slepton mass terms are given by,
\begin{equation}
  - \mathcal{L} \supset
  \wt{d}_{Ai}^\dagger \, [m^2_{\wt{d}AB}]_{ij} \, \wt{d}_{Bj} +
  \wt{e}_{Ai}^\dagger \, [m^2_{\wt{e}AB}]_{ij} \, \wt{e}_{Bj} ,
\end{equation}
in the basis where the down-type quark and the charged lepton mass matrices
are diagonal.
The sfermion mass matrices include contributions from
the Yukawa couplings, the $\mu$ term, the $D$ terms,
the soft scalar mass terms, and the $A$ terms.
In terms of the mass matrices, mass insertion parameters are defined by
\cite{Hall:1985dx}
\begin{equation}
  \begin{aligned}
  \ded{ij}{AB} &\equiv [m^2_{\wt{d}AB}]_{ij} / \msd^2 , \quad
  \del{ij}{AB}  \equiv [m^2_{\wt{e}AB}]_{ij} / \msl^2 , \quad
  (A,i) \neq (B,j) ,
  \\
  \ded{ii}{AA} &\equiv \del{ii}{AA} \equiv 0 ,
  \end{aligned}
\end{equation}
where $A, B = L, R$ denote the chiralities, $i, j = 1, 2, 3$
are the family indices,
and $\msd^2$ and $\msl^2$ are
the average sfermion masses \cite{Gabbiani:1996hi}.
In this work, we heavily rely on the mass insertion notation
defined above to discuss the flavor structure of squarks and sleptons.
Yet, we do not use mass insertion \emph{approximation}
to compute physical amplitudes, but
work with mass eigenstates and mixing matrices.

Normally, as its name implies, a mass insertion is a quantity
that should be defined at the scale of the particle mass.
Therefore, a squark or a slepton mass insertion is considered
at the sparticle mass scale or at the weak scale.
This is the case in the previous paragraph.
In this work, we borrow this notation
to deal with the scalar mass matrices at the GUT scale:
a GUT scale mass insertion is an off-diagonal entry of
a soft scalar mass matrix divided by the averaged diagonal element,
in the basis where the Yukawa matrix is diagonal.
Following this definition, we have
\begin{equation}
  \begin{aligned}
    \ded{ij}{LL} &= [m^2_q]_{ij} / \msd^2, &
    \ded{ij}{RR} &= [m^2_d]_{ij} / \msd^2,
    \\
    \del{ij}{LL} &= [m^2_l]_{ij} / \msl^2, &
    \del{ij}{RR} &= [m^2_e]_{ij} / \msl^2,
  \end{aligned}
\end{equation}
for $i \neq j$.

\end{document}